\documentclass[12pt,a4paper]{article}
\usepackage{float}
\usepackage{amsfonts}
\usepackage{amsmath}
\usepackage{amssymb}
\usepackage{amsthm}
\usepackage{caption}
\usepackage{fontenc}
\usepackage{graphicx}
\usepackage{ucs}
\usepackage[utf8]{inputenc}
\usepackage[left=1.50cm, right=1.50cm, top=1.50cm, bottom=2.00cm]{geometry}
\usepackage{makeidx}
\usepackage{multicol}
\usepackage{pst-all}
\usepackage{rotating}
\usepackage{subfigure}
\usepackage{upgreek}
\usepackage[english]{babel}
\usepackage[]{cite}
\usepackage{xcolor}
\usepackage{cancel}
\usepackage{listings}
\usepackage[breaklinks=true]{hyperref}
\usepackage[normalem]{ulem}

\newcommand {\pythia}       {{\tt PYTHIA8} \cite{Sjostrand:2014zea,Sjostrand:2006za}}

\newcommand {\madgraph}     {{\tt MadGraph\_aMC@NLO} \cite{Alwall:2014hca}}
\newcommand {\mg}           {\tt MadGraph5}

\newcommand {\fastjet}      {{\tt fastjet} \cite{Cacciari2012}}

\newcommand {\heptoptagger} {{\tt HEPTopTagger2} \cite{Plehn:2011tg,Kasieczka:2015jma,Plehn:2010st,Plehn:2009rk,Butterworth:2008iy}}
\newcommand {\toptag}       {\tt HEPTopTagger2}

\newcommand {\feynrules}     {{\tt FeynRules} \cite{Alloul:2013bka}}

\newcommand {\cajet}        {{\tt Cambridge/Aachen} (C/A) \cite{Dokshitzer:1997in}}

\newcommand {\gL} {\ensuremath{g^{\tilde{t}_{1L}}}}
\newcommand {\gR} {\ensuremath{g^{\tilde{t}_{1R}}}}

\definecolor{mred}{rgb}{0.4,0.0,0.0}
\definecolor{mblue}{rgb}{0.0,0.0,0.4}

\definecolor{mGreen}{rgb}{0,0.6,0}
\definecolor{mGray}{rgb}{0.5,0.5,0.5}
\definecolor{mPurple}{rgb}{0.58,0,0.82}
\definecolor{backgroundColour}{rgb}{0.95,0.95,0.92}

\newcommand{\st}{\tilde{t_1}}
\newcommand{\neut}{\tilde\chi^0}

\newcommand{\ctht}{\cos{\theta_{\tilde{t}}}}
\newcommand{\cths}{\cos\theta^{*}}
\newcommand{\be}{\begin{equation}}
\newcommand{\ee}{\end{equation}}
\newcommand{\br}{\begin{eqnarray}}
\newcommand{\er}{\end{eqnarray}}
\newcommand {\secref} [1] {Sec.~\ref{#1}}
\newcommand {\eqnref} [1] {Eq.~\ref{#1}}
\newcommand {\figref} [1] {Fig.~\ref{#1}}
\newcommand{\comment}[1]{}
	
\begin{document}
	\begin{titlepage}
		
		\begin{center}{\Large {\bf Boosted Top quark polarization}}
			\\[8mm]
			
			Rohini Godbole$^{\star}$\footnote{Email: \texttt{rohini@iisc.ac.in}},
            Monoranjan Guchait$^{\dag}$\footnote{Email: \texttt{guchait@tifr.res.in}},
			Charanjit K. Khosa$^{\S}$\footnote{Email: \texttt{charanjit.kaur@sussex.ac.uk}},
            Jayita Lahiri$^{\ddagger}$\footnote{Email: \texttt{jayitalahiri@hri.res.in}},
            Seema Sharma$^{\P}$\footnote{Email: \texttt{seema@iiserpune.ac.in}} and
            Aravind H. Vijay$^{\dag}$\footnote{Email: \texttt{aravind.vijay@tifr.res.in}}
		\end{center}
		
		\vspace*{0.20cm}
		
		\centerline{$^{\star}$ \it Centre for High Energy Physics, Indian Institute of Science, Bangalore-560012, India}
		\vspace*{0.4cm}
		\centerline{$^{\dag}$ \it Department of High Energy Physics, Tata Institute of Fundamental Research,}
		\centerline{\it Homi Bhabha Road, Mumbai-400005, India}
		
		\vspace*{0.4cm}
		
		\centerline{$^{\S}$ \it Department of Physics and Astronomy, University of Sussex,}
		\centerline{\it Brighton, BN1 9RH, United Kingdom}
		
		\vspace*{0.4cm}
		
		\centerline{$^{\ddagger}$ \it Regional Centre for Accelerator-based Particle Physics,}
		\centerline{\it Harish-Chandra Research Institute, HBNI, Chhatnag Road, Jhunsi, Allahabad - 211 019, India}
		
		\vspace*{0.4cm}
		\centerline{$^{\P}$ \it Indian Institute of Science Education and Research, Pune-411008, India}
		
		\vspace*{1.2cm}
		
		\begin{abstract}
            In top quark production, the polarization of top quarks, decided by the chiral structure of couplings, is likely to be modified in the presence of any new physics contribution to the production.
            Hence it is a good discriminator for those new physics models wherein the couplings have a chiral structure different than that in the Standard Model (SM).
            In this note we construct probes of the polarization of a top quark decaying hadronically, using easily accessible kinematic variables such as the energy fraction or angular correlations of the decay products.
            Tagging the boosted top quark using 
            jet sub structure technique, we study robustness of these observables for a benchmark process, $W^{\prime} \to tb$.
            We demonstrate that the  energy fraction of b-jet in the laboratory frame and a new angular variable, constructed by us  in the top rest frame, are both very powerful tools to discriminate between the left and right polarized top quarks.
            Based on the polarization sensitive angular variables, we construct asymmetries which reflect the polarization. We study the efficacy of these variables for two new physics processes where which give rise to boosted
            top quarks:
            (i) decay of the top squark in the context of supersymmetry searches, and
            (ii) decays of the Kaluza-Klein(KK) graviton and KK gluon, in Randall Sundrum(RS) model.
            Remarkably, it is found that the asymmetry can vary over a wide range about
            +20\% to -20\%. The dependence of asymmetry on top quark couplings of the new particles present in these models beyond the SM (BSM) is also investigated in detail.
		\end{abstract}
	\end{titlepage}

\section{Introduction}
\label{sec:intro}
Top quark is an interesting object in the standard model (SM),
since it is the heaviest known fermion and it has the
strongest coupling with the Higgs boson, close to unity.
It decays before hadronization, and  hence soft QCD effects do not wash out its spin information and
the decay product kinematic distributions reflect the same.
Because of its large  mass, top quark 
can be closely related to the phenomenon
of  electroweak symmetry breaking.
This may be either in the context
of the  Higgs model or any alternative mechanism by
which fundamental particles acquire masses.
Therefore, understanding the properties of the top quark,
specifically the Lorentz structure of
its couplings, responsible both for the production and decay
has  always received special attention
\cite{Beneke:2000hk,Husemann:2017eka},
and holds the potential  of offering us a glimpse of BSM.
To this end, an important and interesting property is its polarization which
is reflected in the the kinematics of its decay products.

In hadron colliders, top quarks are produced as $t \bar{t}$ pairs via
strong interaction or in association with a $b$ quark or $W$ boson
via electroweak interaction.
Note that in case of pair production, the top quarks are
mostly unpolarized,  due to the vector nature of the dominant QCD couplings,
although their spins are correlated. These spin-spin correlations are studied quite well, both theoretically and experimentally
\cite{Barger:1988jj,Uwer:2004vp,Mahlon:1995zn,
    Baumgart:2012ay,Aaltonen:2010nz,Khachatryan:2016xws,
    Aad:2014mfk}.
On the other hand, the produced single top quark is
polarized, and in SM it is purely left chiral due to
the V-A nature  of  the $t-b-W$ interaction.
Any modification of the tensor structure of this interaction
at the production vertex deviating from the SM
would change the polarization of produced top quark.    
Hence, the measurement of polarization of top quark
is expected to be a good probe
of new physics in interactions responsible for the production of the top quark.
Angular distribution of decay products,
in particular the visible lepton from its semileptonic
decay is found to carry polarization information of the
parent top quark \cite{Jezabek:1994qs}.
Extraction of polarization information of top quark
by exploiting the various kinematic
distributions of charged lepton is discussed
in great detail in the literature
\cite{Jezabek:1994qs,Grzadkowski:2001tq,Grzadkowski:2002gt,
 Godbole:2006tq,Godbole:2007zzb,Godbole:2010kr,Hagiwara:2017ban,Jueid:2018wnj}. 
Moreover, top polarization also has been studied well using the 
matrix element method\cite{Brandenburg:2002xr,Baumgart:2012ay,Tweedie:2014yda,Abazov:2011gi}

In this study we attempt to develop a strategy to measure the top quark polarization in its hadronic decay, in particular when it is boosted. 
The main motivation of this study is in the context of new physics searches 
in scenario where decay of a heavy particle results in a top quark which is boosted, 
and in principle also polarized. To be specific, we consider 
few new physics processes, where polarized top quarks originate from 
heavy particles such as $W^\prime \to t b$, the superpartner of top, 
top squark($\tilde{t}$), and Kaluza-Klein states in extra dimension models.
The polarization of the produced top depends on the chiral structure of the 
respective couplings responsible for the decay of the new physics particle.
In view of the current exclusion limits on these particles, it is clear that
the top quarks produced in the decay would be highly boosted.
All the decay products from boosted top quarks emerge
in a single cone along the direction of top quark without
much angular separation, which makes it difficult
to construct clean polarization sensitive observables.
Thus  measurement of polarization of boosted top
quark though  a challenging task, is of great importance for developing a tool
to probe  the nature of top couplings with new particles  in the context of 
new physics searches.
Ideally, as we mentioned already,
semi-leptonic decay mode of the top quark is very well suited for studying the spin effects, as the
charged lepton with the largest spin
analyzing power is easy to identify,
and is not affected much by soft QCD radiation
\cite{Jezabek:1994qs}.
Equally importantly, angular distribution of the lepton with respect to the spin direction for a polarized top quark is unchanged by anomalous $tbW$ couplings, to linear order
\cite{Godbole:2018wfy,Grzadkowski:2001tq,Grzadkowski:2002gt,Godbole:2006tq,Godbole:2007zzb}
and hence is a good probe of top polarization unaffected by new physics in decay.
The price to pay for the leptonic decay is low branching
ratio, and difficulties in finding an isolated
lepton originated from boosted top.
In addition, presence of multiple sources of missing
energy makes it difficult to reconstruct the top quark
momentum. Reconstruction of
top quark momentum for
leptonic decay, even in the presence of many sources
of missing energy is likely to be feasible but needs
a very non trivial algorithm \cite{Plehn:2011tf}.

The above arguments suggest that the hadronic decay mode of top quark  
is more favorable one to reconstruct the boosted top quark,
and study its spin effect where the down-type quark from $W$ decay
plays the same role as the charged lepton.
It is to be noted that the reconstruction of jets and
its identification corresponding to the
down-type quark from $W$ decay involves certain
level of uncertainties.
The decay products of boosted top quark in its hadronic
channel are identified by employing the powerful technique
of jet substructure analysis~\cite{Butterworth:2008iy}.
In this paper, we discuss how the polarization of boosted 
top quark in its hadronic decay mode can be measured
by constructing the polarization sensitive observables out
of the momenta of subjets inside the reconstructed top jet.
Polarization of boosted tops using jets at the
substructure level has been discussed in the literature
\cite{Tweedie:2014yda,Shelton:2008nq,
	Kitadono:2015nxf,Conway:2016caq,Lapsien:2016zor,
	Krohn:2009wm,Bhattacherjee:2012ir,Kitadono:2014hna}.

For hadronically decaying polarized top quarks, down-type quark from $W$ 
decay with the largest spin
analyzing power, is mostly softer than the other two quarks in the top 
rest frame as its momentum is opposite to the boost direction of the 
$W$ due to the angular correlations.
Certainly, in this analysis, the main challenge is
to retrieve the identity of this quark as accurately as possible.
Many interesting proposals, in this context, are summarized in \cite{Tweedie:2014yda}.
Notably, in this paper the author proposes
a new axis taken as the
weighted average of the two jet axis, where the weights
are given by the probability obtained from the decay
matrix element.
Interestingly, there are observables\cite{Krohn:2009wm} which
could discriminate left and right polarized
boosted hadronically decaying top quarks without requiring $W$ reconstruction
or $b$ jet identification.
In this case,  subjets with good spin analyzing power
are identified by
requiring the corresponding pair of subjets with the
minimum $k_T$ distance.
In our study, focusing on the single-top polarization effects,
first, we revisit some of these observables, namely energy fraction
of top quark decay products, proposed in earlier studies
\cite{Krohn:2009wm}, and their role as a polarimeter.
Currently, tagging of high $p_T$ $b$-jets using MVA
techniques with a reasonable efficiency is not difficult as shown
by experimental studies \cite{CMS:2016kkf,Aaboud:2018xwy}.
Therefore, we employ energy fraction of tagged high $p_T$ $b$ jets as
one of the useful polarimeter \cite{Roy:2018nwc}.
In addition to these energy fraction observables, we also propose
a robust polarization sensitive variable related indirectly 
to the angular distribution of the
selected subjet corresponding to the down-type quark from W decay.
Experimentally it is not very difficult to measure 
this variable event by event. Exploiting this observable, 
an asymmetry of event can be predicted, which may reflect the 
polarization of the decaying top unambiguously. 
We demonstrate the impact of the polarization sensitive
observables discussed above by choosing the following benchmark process,
\be
{pp} {\rightarrow} {W^{\prime}} {\rightarrow} {tb} 
\label{eq:wprime}
\ee
where the $W'$ gauge boson couples with both left
and right handed top quark which is assumed to decay,
$t \to b W$ with 100\% branching ratio.
We simulate the process of Eq.~\ref{eq:wprime} and study  
the observables  which have a potential to discriminate between 
left and right polarized top quarks.
After demonstrating efficacy of these variables as a 
polarimeter for completely polarized top quarks, we
use them in situations where the top polarization has a value 
different from $\pm 1$.
For example, top squark,
produced in pairs at the LHC decay to top quark accompanied by the
lightest neutralino($\neut_1$), $\st \to t \neut_1$
~\cite{Aad:2014mfk}.
The polarization of this top quark is controlled
by the $\st - t - \neut_1$ coupling, which has both
gauge and Yukawa type of interactions.
A detailed study shows how the polarization of top quark
from top squark decay can be identified by measuring the asymmetry
of events.
The dependence of this asymmetry and its sensitivity to
the variation of this coupling strength is investigated.
This study reveals how the polarization of top quark possibly can
shed some light about the 
 $\st - t - \neut_1$
couplings in supersymmetric(SUSY) theories.
Similar studies are also carried out 
in the process where polarized top quarks are produced from the decay of
Kaluza-Klein(KK) excited states.

The paper is organized as follows. In \secref{sec:form},
we briefly discuss  the kinematics of decay products in the
context of polarized top decays, Then 
in \secref{sec:BoostedTopPol} we identify observables which can be used
to  obtain information about the polarization of
top quark.     
In \secref{sec:MSSM}, we study the application 
of the proposed variables in top squark decay which can 
produce top quarks which are not necessarily completely polarized.  In 
\secref{sec:RSmodel}, similar studies are carried out 
in the context of Randall Sundrum(RS) model where polarized top
quarks are produced in the decay of new resonances. Finally in \secref{sec:summary}
we summarize our results.

\section{Polarized top quark decays}
\label{sec:form}
It is instructive to review briefly the effect of top
polarization on the kinematics of its decay
products, before we proceed further to discuss our results.
In the SM, the top quark has $\sim70\%$ branching fraction for the  hadronic
channel, $t \to b~W^+ \to b~u~\bar d$, where $u(\bar{d})-$quark
represents the up(down) type quarks.
The angular distribution of any of the decay products 
from the top quark (in the top rest frame)
can be written as \cite{Jezabek:1988ja},
\begin{eqnarray}
\frac{1}{\Gamma}\frac{d\Gamma}{d\cos\theta_{f}} &
= &
\frac{1}{2}\left(1+{\cal P}_{0}\kappa_{f}\cos\theta_{f}\right)
\label{eq:DecayDist}
\end{eqnarray}
where $\kappa_{f}$ is the spin analyzing power of the respective decay particles, i.e. $f=b$, $\bar d$, $u$ and $W$.
${\cal P}_{0}$ is the polarization of the decaying top~\cite{Mahlon:1999gz,Schwienhorst:2010je,Shelton:2008nq,V.:2016wba}
$(-1\le {\cal P}_{0} \le1)$
Here $\theta_f$ is the angle between the fermion (or $W$) and
top spin direction, in the top quark rest frame.
The spin analyzing power $\kappa_f$, can
be calculated, and its values are given by
$\kappa_{\bar d} = 1$, $\kappa_u\approx-0.3$ and
$\kappa_{b}\approx-0.4$ at tree level in the SM
\cite{Jezabek:1994qs}.
Notice that when top decays hadronically, the down-type quark($\bar d$)
has the maximum spin analyzing power, i.e it is strongly
correlated with the top spin.
However, the b quark is also a good candidate with $\kappa_{b}\approx-0.4$, to study
the top polarization.
In principle,
the reconstructed
$W$, on the other hand, with opposite spin analyzing power of $b$ quark ($k_W \approx 0.4$), in
the top quark rest frame can also serve as a good top spin analyzer.
Therefore, the kinematic distribution of down-type quark or b-jet(or $W$)
is expected to provide considerable handle to study top polarization,
and we take the spin direction of the
top to be quantized along its momentum direction.
The corrections to the tree level values of spin analyzing power
of hadronic decay products are found to be at the level of 3-4\%  
\cite{Brandenburg:2002xr,Bernreuther:2014dla}.

As pointed out earlier, identification of subjet
corresponding to ${\bar{d}} { \left( d \right) }$ quark
from $W^+(W^-)$ decay having maximum spin analyzing power
is non trivial, specially in the busy environment of
hadron collider, and in particular when the three jets are not
well separated because of large boost of the parent top quark.
To this end, we try to identify the subjet with
large spin analyzing power using the following
methods:
\begin{enumerate}
    \item
    The subjet which is aligned along the $b$-like jet, i.e. constitutes
    a minimum invariant mass with the $b$-jet.
    
    \item
    The harder of the two subjets which have the minimum $k_T$ distance between each other \cite{Krohn:2009wm}.
    
    \item
    In addition, $b$-jet can be used as a good candidate to study
    top polarization, provided it can be tagged even with high momentum.
    Currently, techniques are developed using MVA based methods to
    tag $b$-jets of high $p_T$
    \cite{1748-0221-13-05-P05011,ATLAS-CONF-2016-039}.
    
\end{enumerate}
Furthermore, in this context we propose a new observable indirectly related to
the angular distribution of the decay products of a top quark, 
which turns out to be very robust in measuring the top polarization.
To construct this observable, the b-like subjet inside the tagged top 
jet is identified after
reconstructing the $W$ mass out of three subjets.
We label the constituent subjets of reconstructed
$W$ as $j_1$ and $j_2$ such that $m_{bj_1} < m_{b j_2}$.
Indeed about $\approx 50 \%$ to $60 \%$ \cite{Brandenburg:2002xr,Tweedie:2014yda}
of cases $j_1$ will act as a proxy
for the parton-level $d$ quark, which has the maximum spin-analyzing power.
The momentum direction of $j_1$
is guided by the
polarization of top quark.
In the lab frame, the direction of momentum and the spin of top jet 
is opposite to each other 
for left handed, and in the same direction for the right handed top quark.
Hence the angular correlation 
between the momenta of $j_1$ sub jet and topjet is
expected to show effect of polarization of the parent top quark.
With this understanding, we construct the observable,
namely $\cos\theta^\star$, as defined,
\begin{eqnarray}
\cos \theta^{*} & \equiv &
\frac{ {{\vec{ t_j} }} \cdot {{\vec{j}_1}^{\prime}} }
{ \left| {{\vec{t_j}}} \right|\left| {{\vec{j}_1}^{\prime}} \right|},
\label{eq:ourobs}
\end{eqnarray}
where ${\vec{t}_j}$ is the momentum of reconstructed 
top jet in the lab frame, and 
${\vec{j}_1}^\prime$ is the sub jet momentum in  top jet rest frame.  
Essentially, this variable is defining the direction of the momentum of 
the subjet $j_1$ in the top rest frame, with respect to the direction of
top jet momentum in the lab frame.  
Hence, identifying the momentum of the 
subjet $j_1$, following the above method  the
angular variable $\cos \theta^*$ can be computed easily.
In the next section,
we demonstrate the robustness of $\cos\theta^\star$ along with
other kinematics observables as presented above, considering
a benchmark process~\eqnref{eq:wprime}, 

\section{Boosted top polarization}
\label{sec:BoostedTopPol}
The top quark produced through the decay of a heavy BSM particle, 
(\eqnref{eq:wprime}), is expected to be
boosted. Hence the decay products of top will be 
collimated, and form a single fat jet of large cone size.
Identification of boosted objects using jet substructure
technique is now very well established and tested
\cite{Plehn:2011tg,Kasieczka:2015jma,Plehn:2010st,
	Plehn:2009rk,Butterworth:2008iy,Kaplan:2008ie}.
In this technique,
topjets are tagged  by finding the substructures of fat jets 
following BDRS mass drop \cite{Butterworth:2008iy} and
filtering method using {\heptoptagger}.

The matrix element for the benchmark process, \eqnref{eq:wprime},
is generated using the {\feynrules} corresponding to the
$W^{\prime}$ effective model \cite{Fuks:2017vtl}
in {\madgraph} at $\sqrt{s}=13$ TeV to generate left
handed and right handed top quark via s-channel.
It is to be noted that the whole process up to the top decays to jets are
treated in matrix element and hence the production and decay are not
factored separately, which ensures accurate treatment of finite widths
and interference terms with the spin effects.
The partonic events are then showered using {\pythia}.
This model is an extension of the SM, including an additional interaction
of fermions to $W^{\prime}$ boson following the lowest-order effective 
lagrangian, described in Refs \cite{Sullivan:2002jt,Duffty:2012rf},
\be
	{\cal L} = \frac{V_{ij}}{2\sqrt{2}} \bar{f_i}\gamma^{\mu}
	\left({g_{R}}(1+\gamma_5)+{g_{L}}(1-\gamma_5)\right)W'_{\mu}f_j
	+ h.c.
	\label{lagwprime}
\ee
It is clear from the above equation that the coupling
strengths, $g_{R}$ and $g_{L}$
decide the polarization of the produced top quark in $W^{\prime}$ decay. 
For instance, if $g_{L} = 1$, and $g_{R} = 0$,
the produced top quark will be left-chiral and if $g_{L} = 0$,
and $g_{R} = 1$, it will be right-chiral. For the case, $m_{W^{\prime}} \gg m_t$,
the produced top is highly boosted and hence its helicity (polarization)
is almost same as it's chirality.
In the simulation, fat jets are reconstructed using {\cajet} 
algorithm setting the jet size parameter $R=1$, and required to 
have transverse momentum $p_T > 200$ GeV 
and $\left| \eta \right| < 4$.
Generated events
consisting of fat jets are passed
through {\toptag} to tag top jet by BDRS mass drop method
using default set of parameters for mass cuts and filtering.
The correspondence of the parent top quark with the reconstructed top
jet is ensured by matching with cone size $\Delta R=0.3$.
As mentioned in earlier sections, the momentum distribution
of subjets of boosted top 
quark are guided by its state of polarization.   
This fact leads to a difference in the tagging
efficiency for left and right handed top quark as also reported by ATLAS
collaboration~\cite{Aaboud:2018juj}.
\begin{figure}[h]
	\begin{center}
		\includegraphics[width=0.5\textwidth]
		{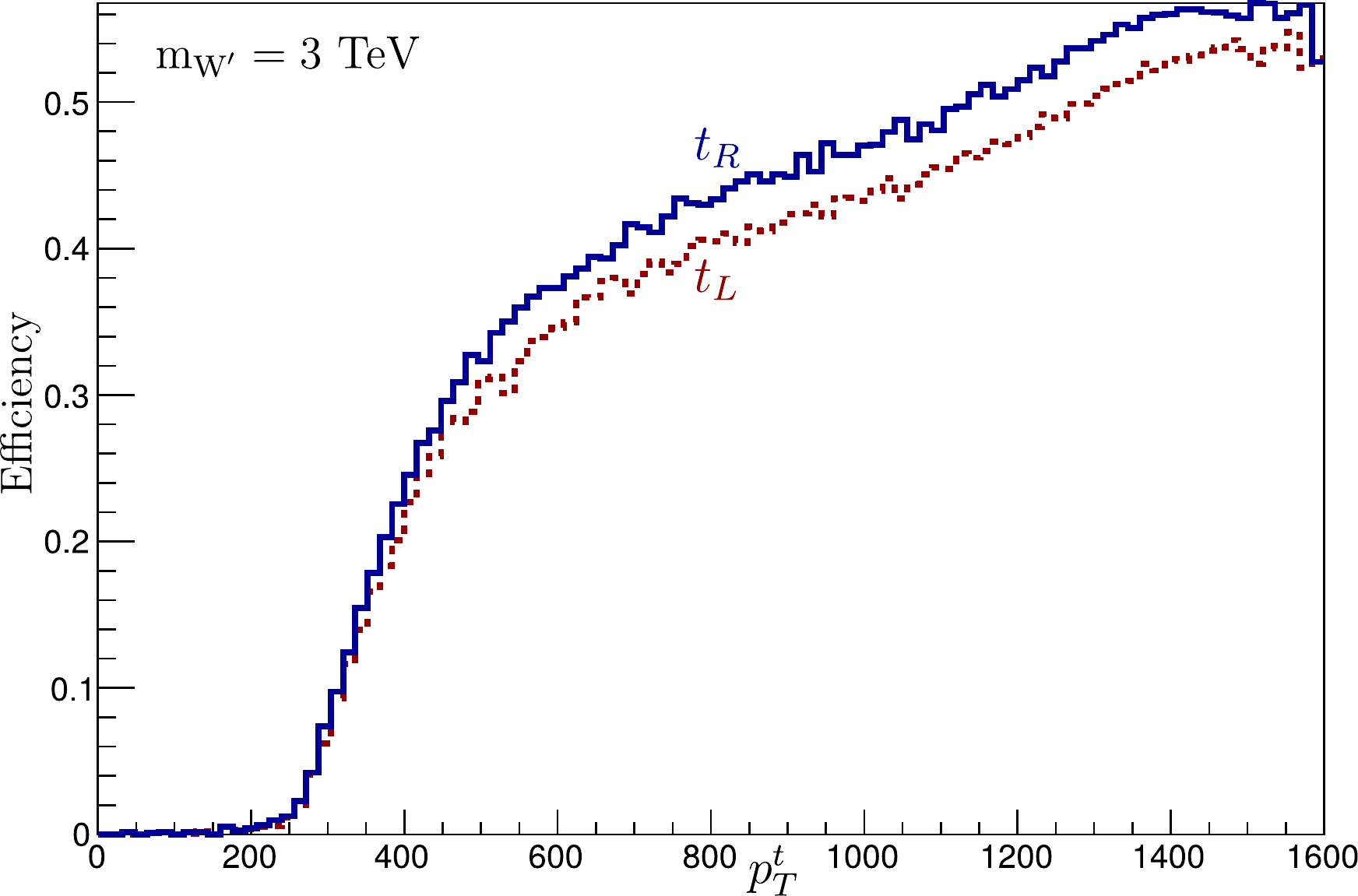}
		\caption{
			Top tagging efficiency for left ($t_L$) and right 
			chiral ($t_R$) top quark for the process $ p p \rightarrow W' \rightarrow t b$ 
			with  $m_{W^\prime}=3$~TeV.
		}
		\label{fig:TaggingEfficiency}
	\end{center}
\end{figure}
We also observed this difference as shown
in \figref{fig:TaggingEfficiency}, the top tagging efficiency
for both left handed and right handed cases
as a function of $p_T$ of the parent top quark.

As can be seen from \eqnref{eq:DecayDist}, and the following discussion
on spin analyzing power, in case of right handed top quarks (in the lab frame),
the $\bar{d}$ quark is more boosted while the other two quarks
($b, u$) are less boosted, and hence more separated
as compared to left handed tops where $b, u$ quarks are more boosted
and hence less separated.
It implies that in case of right handed top quarks, the subjets are better separated
than the case for
left handed top quarks.
This difference of kinematics
for left handed and right handed tops leads to a small difference
in top tagging efficiency as seen from \figref{fig:TaggingEfficiency}.

\begin{figure}[H]
	\begin{center}
		\includegraphics[width=0.49\textwidth]
		{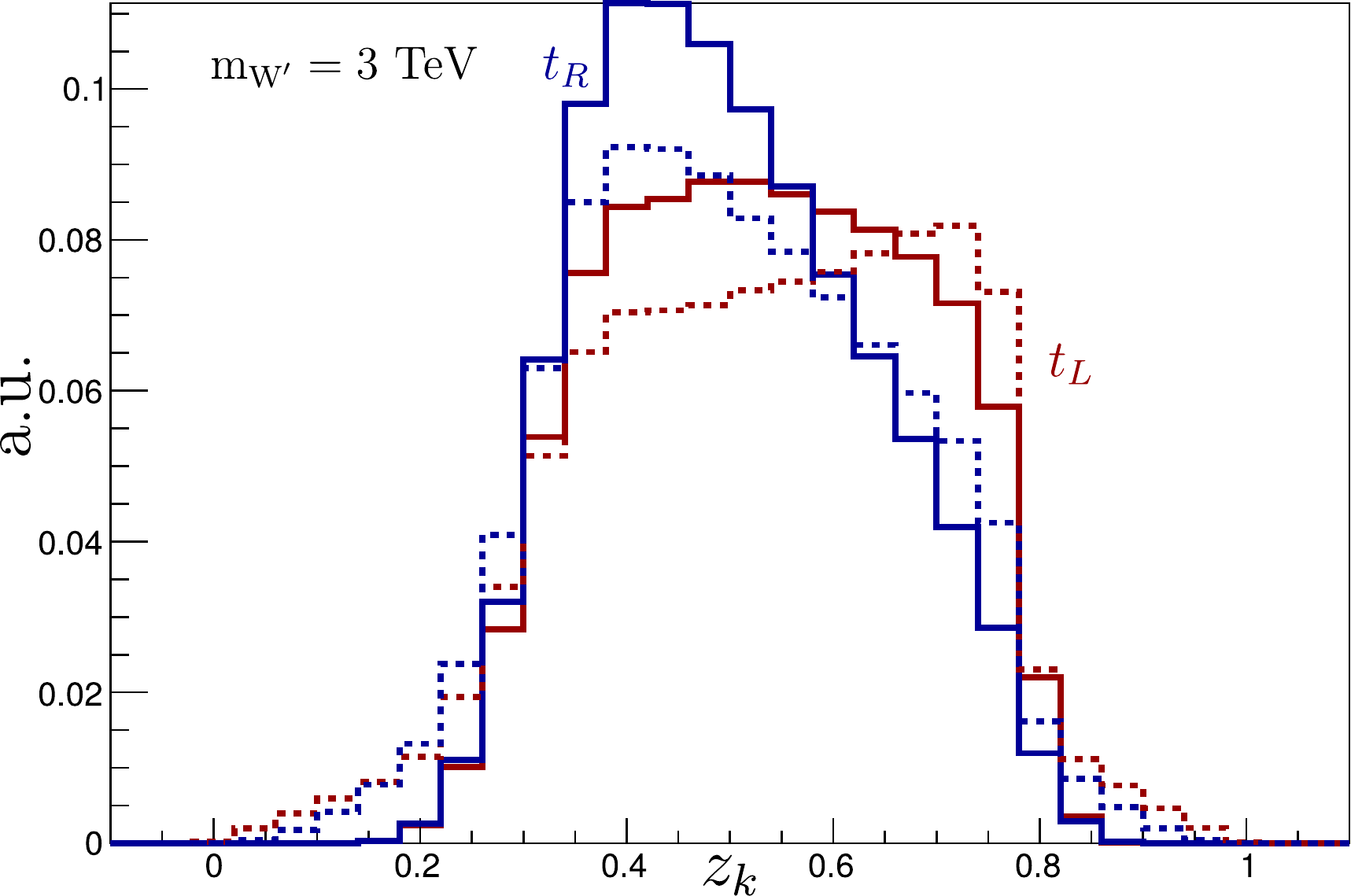}
		\caption{
			The energy fraction $z_k$ (\eqnref{eq:kronzk}) 
for left (red) and right handed 
			(blue) top quark
			for $m_{W^{\prime}}=3$ TeV.
			The dashed lines correspond to
            the observable at the partonic level.
		}
		\label{fig:Krohn_Compare}
	\end{center}
\end{figure}

Now we present the kinematic observables
discussed in the previous section, which could be used to
distinguish the left and right handed boosted top.
As a first step, we study the energy fraction
variable($z_k$) which is also suggested in Ref.~\cite{Krohn:2009wm}.
As stated before, the jet corresponding to $\bar{d}$ is relatively harder 
for right handed (helicity) top and b-like jet is 
harder when it is left handed (helicity). With this understanding 
we define and then examine the $z_k$ variable.
The tagged topjet corresponding to top quark
contains at least three subjets.
Among the three possible combinations, the pair of subjets having the 
smallest $k_T$
distance is identified. The $k_T$ distance $d_{ij}$ is defined as,
\be
d_{ij}=\min(p_{Ti}^2,p_{Tj}^2)R_{ij}^2,
\ee
where $R_{ij}^2 = (y_i-y_j)^2 + (\phi_i-\phi_j)^2$.
The energy
fraction of the harder jet $j_k$ of this pair
having smallest $k_T$
is defined as $z_k$
\cite{Krohn:2009wm}, i.e.,
\begin{eqnarray}
    z_{k} & = & \frac{\max\left(E_{i},E_{j}\right)}{E_{t}}
    \text{, where } d_{ij} \text{ is minimum.}
    \label{eq:kronzk}\\
    \text{and }E_t& : & \text{energy of the top jet,} \nonumber
\end{eqnarray}
acts as a good polarimeter \cite{Krohn:2009wm}.
For the sake of comparison, in order to understand how the
$z_k$ works, we also compute it using the partonic(truth)
level information.
Both the results are presented in \figref{fig:Krohn_Compare}
where the energy fraction $z_k$ is shown (solid) along with
the truth level distribution(dashed) for both left and right
handed top quark originating from $W'$ decay.
As expected, it is seen that by matching subjets 
with the parton level quarks, about 50\% of the cases, algorithm chooses
b-like subjets for the left-handed top quarks case and
d-like subjets for the right-handed top quarks.

As mentioned in the previous section that
the $\bar d$ quark is maximally correlated with the top spin, hence
for left-handed top (in the lab frame), the $\bar{d}$ quark
tends to be soft. Thus the minimum $k_T$ pair tends
to involve the $\bar d$-like subjet
and the other one mostly $b$-like with comparatively
harder energy.
For right handed top quark case, we observe mostly 
the same kind of pairing, but with harder $\bar d$-like
subjet and softer $b$ like subjet.
Notice that the shape of $z_k$ distribution is not so different
from partonic level at the intermediate values of
$z_k \approx 0.5$.
Thus the energy fraction variable $z_k$ can play the role of a
polarimeter in differentiating the left and right handed top quark.

We present another observable, the energy fraction of
$b$-like subjet, defined as follows:
\be
Z_b = \frac{E_b}{E_t}
\label{eq:Zk}
\ee
Distributions of this variable ($Z_b$)
are presented in \figref{fig:BEFrac} for both
left and right handed top quarks.
In our simulation, $b$-like subjets are identified through matching with
the parton level $b$-quark.
Recently, techniques are developed to tag 
$b$-subjets inside boosted tagged top jets.
The tagging efficiency is found
to be around $\sim 50-70\%$ depending on background rejection
\cite{1748-0221-13-05-P05011,ATLAS-CONF-2016-039}.
For left handed top quark, the $b$-jet carries
most of the energy of the top jet, whereas for the right handed case, it carries relatively
less energy.       
Comparing the distribution between $z_k$(\figref{fig:Krohn_Compare}) 
and $Z_b$(\figref{fig:BEFrac}), it is found that
$Z_b$ provides a better distinction between left and right handed top quark.

The third observable which we already discussed
in this context is
$\cos\theta^\star$(\eqnref{eq:ourobs}). The distribution
of this angular observable for left and right-handed top quark
as shown in
\figref{fig:cpsthetastar} shows  
remarkable correlation with the polarization of the parent top quark.
The ${\vec t_j}$ (as defined in the previous section)
and top-spin are in the opposite direction at the lab frame
in case of left handed top, and are in same direction
for right handed top.
On the other hand, the momentum direction of the sub jet $j_1$ 
having maximum spin analyzing power ($\bar{d}$-type quark), and with
{comparatively less energy (in the top rest frame)},
is opposite to the spin direction of the top quark.
This argument explains why $\cos\theta^\star$ is
preferred to be positive for right handed and negative for left 
handed top quarks, as shown in \figref{fig:cpsthetastar}.
The dotted lines in this figure represent the same quantity constructed 
out of the parton level momenta of the top decay products. 
As we can see, although the reconstructed distributions are smeared 
but the correlation of the distribution with polarization 
is maintained.
By default, the proposed variables
\eqnref{eq:ourobs}, \ref{eq:kronzk} and \ref{eq:Zk}
are sensitive to the parton showering model used in event generators.
However, these effects are expected to be very small since
these variables are constructed as a ratio of two momenta.

This distinct characteristic of $\cos\theta^\star$ is exploited in
constructing the
event asymmetry defined as,
\begin{eqnarray}
A_{\theta^{*}} & \equiv & \frac{N_{\cos\theta^{*}>0}-N_{\cos\theta^{*}<0}}{N_{\cos\theta^{*}>0}+N_{\cos\theta^{*}<0}}\label{eq:asym}
\end{eqnarray}
where $N$ represents the number of events for the given condition of
$\cos\theta^\star$, either positive or negative.
\begin{minipage}{0.48\textwidth}
    \begin{figure}[H]
        \includegraphics[width=1.0\textwidth]
        {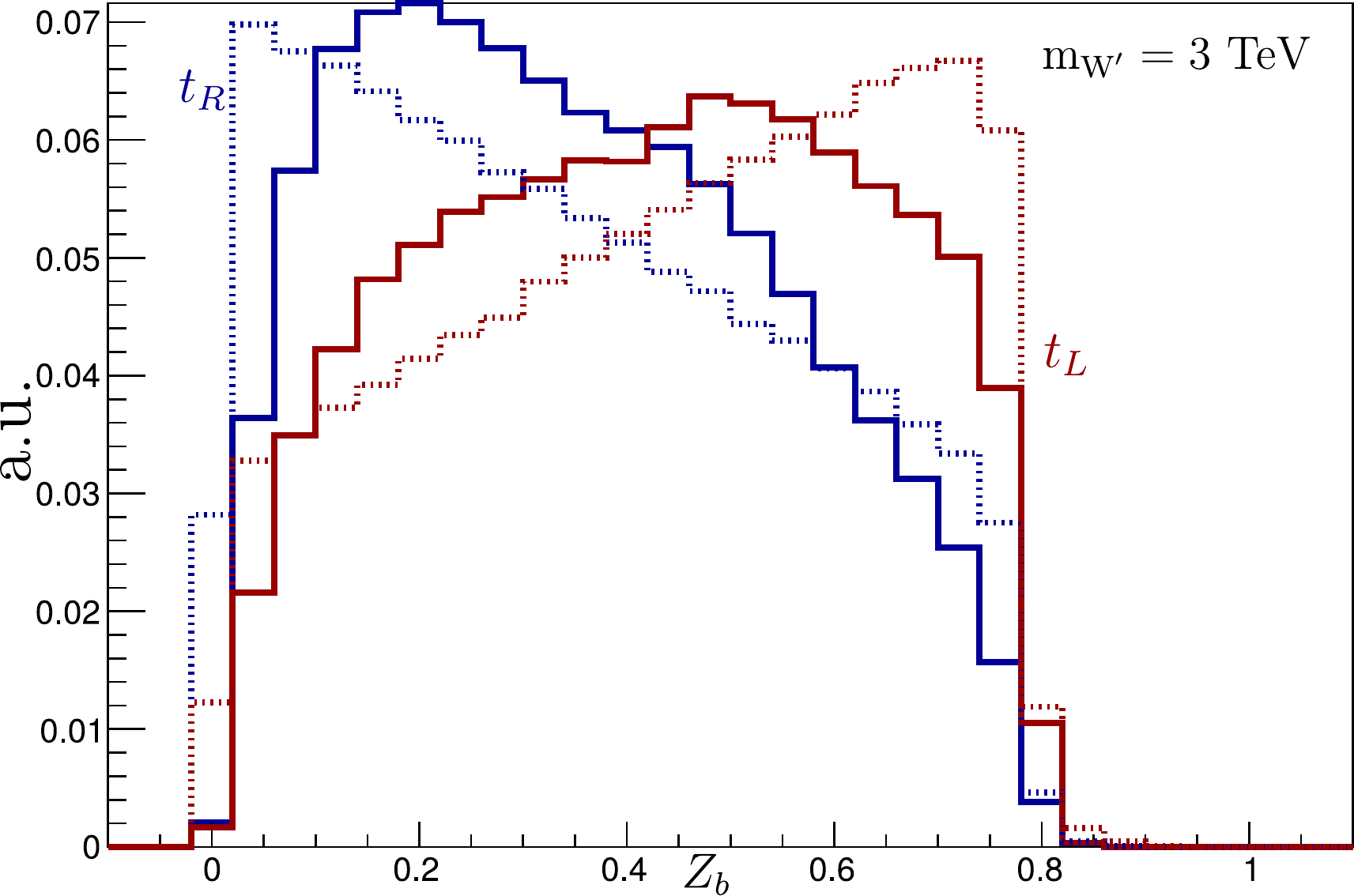}
        \caption{
            Energy fraction of $b$-like subjet for
            partonic (dashed) and reconstructed top jet using {\toptag} (solid).
        }
        \label{fig:BEFrac}
    \end{figure}
\end{minipage} \hspace{0.02\textwidth}
\begin{minipage}{0.48\textwidth}
    \begin{figure}[H]
        \includegraphics[width=1.0\textwidth]
        {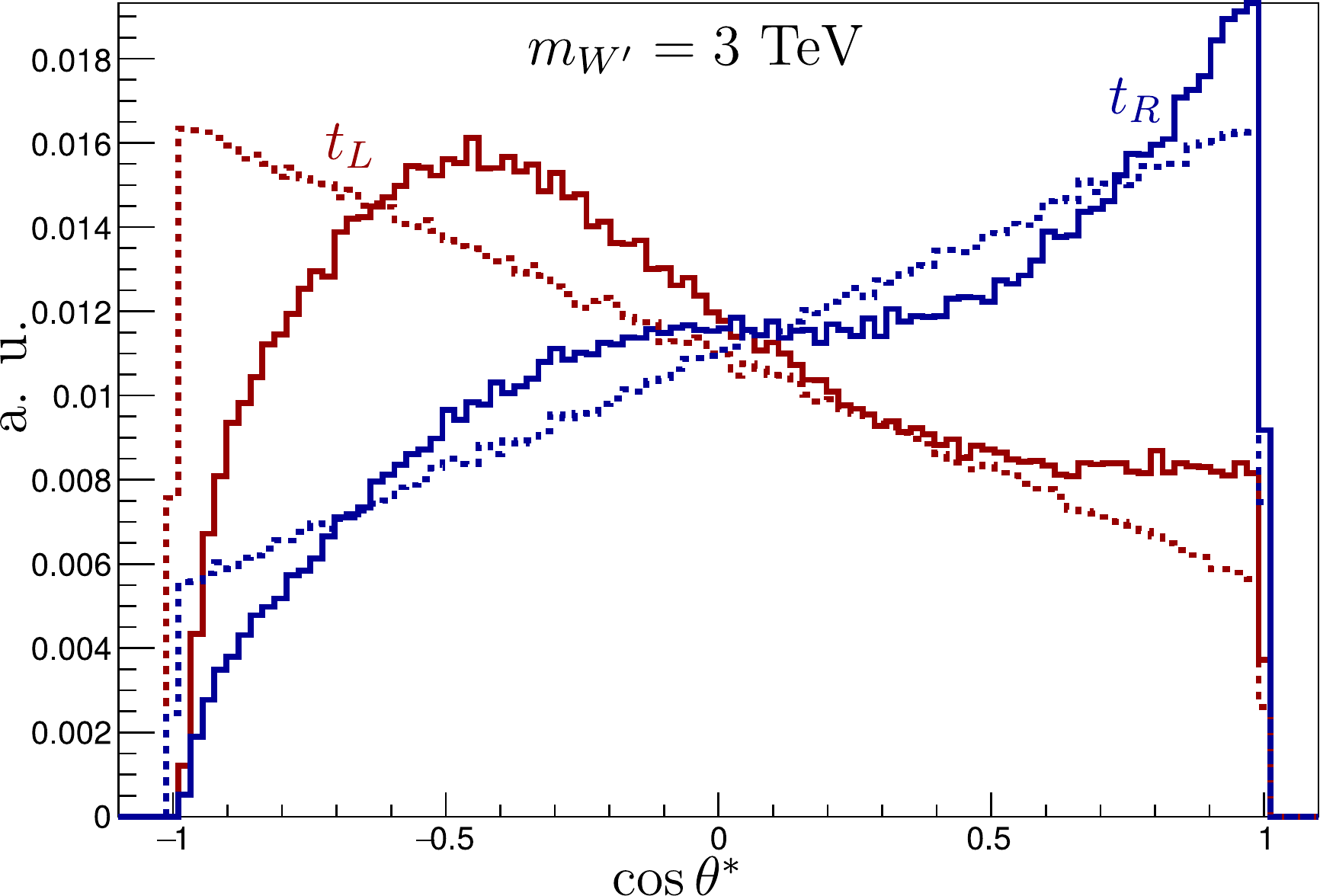}
        \caption{
            The $\cos\theta^*$ distribution for $m_{W^{\prime}}=3$ TeV
            for reconstructed top jet (solid) and partonic (dashed).
        }
        \label{fig:cpsthetastar}
    \end{figure}
\end{minipage}
\begin{figure}[H]
    \begin{center}
        \includegraphics[width=0.5\textwidth]
        {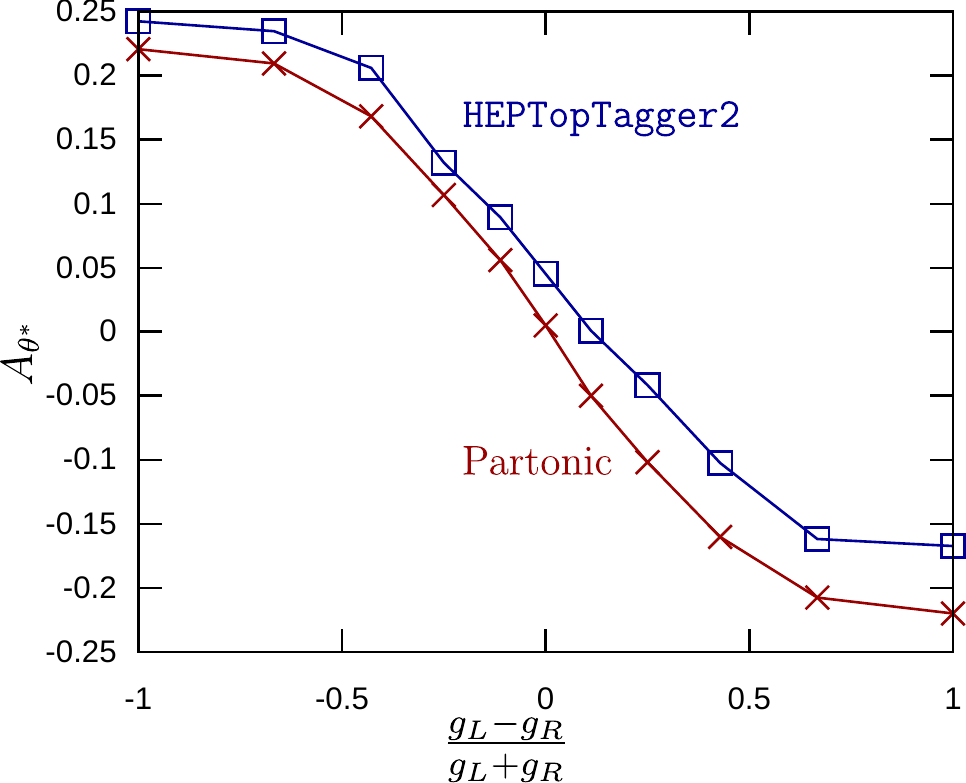}
        \caption{
            The asymmetry ({\eqnref{eq:asym}}) for parton
            level and tagged tops for $m_{W^{\prime}} = 3$ TeV.
        }
        \label{asymwp}
    \end{center}    
\end{figure}
The asymmetry defined above is expected to be very sensitive to the 
type of couplings(see \eqnref{lagwprime}), which decide the polarization
of the top quark in the process, \eqnref{eq:wprime}.
The behavior of it with the combination of 
couplings $\left(g_L - g_R\right)/\left(g_L+g_R\right)$ is 
presented in \figref{asymwp} along with the parton level distribution
where four momenta of quarks are used.
One can clearly see that in the left-handed case i.e. $g_L = 1$ and 
$g_R = 0$, the observed asymmetry is negative, since dominantly 
right handed top quarks are produced, where as it is positive for
the opposite case i.e. for $g_L=0, g_R=1$.
\section{Top polarization in stop decay}
\label{sec:MSSM}
In this section we demonstrate the impact of constructed
discriminating observables
\eqnref{eq:ourobs}, \ref{eq:kronzk} and \ref{eq:Zk}
in distinguishing the left and right polarized top
quarks originating from top squark decays.
Top squark, a colored sparticle can be produced at
the LHC in proton-proton collision,
and subsequently decays in a
variety of channels depending on its mass and composition,
leading to many diverse signatures.
We focus on the decay channel of top squark,
\br
\st \to t \neut_1,
\label{eqn:stopdecay}
\er
which can be the dominant mode with large branching ratio
for a certain region of parameter space where the neutralino
$\neut_1$ is assumed to be the lightest supersymmetric particle.
The coupling between $\tilde{t}_{1}$, $t$ and
$\neut_1$ at the tree level is given by,
\br
{\cal L} = \bar{\tilde{\chi}}_1^0 (\gL P_L + \gR P_R)t \tilde t_1 + h.c,
\er
determines the polarization of top quark
in the final state.
The electroweak correction to this decay channel is at
the level of few percent \cite{Djouadi:1996wt}.
The form of couplings $\gL$ and $\gR$ are,
\br
g^{\st_L} & = & -\sqrt{2}g_{2}
\left[\frac{1}{2}Z_{12}^{*}+\frac{1}{6}\tan\left(\theta_{W}\right)Z_{11}^{*}\right]
\cos\theta_{\tilde t}
-\left[\frac{g_{2}m_{{t}}Z_{14}^{*}}{\sqrt{2}m_{W}\sin\left(\beta\right)}\right]
\sin\theta_{\tilde t}\\
g^{\st_R} & = &
\left[\frac{2\sqrt{2}}{3}g_{2}\tan\left(\theta_{W} \right)Z_{11}\right]
\sin\theta_{\tilde t}-\frac{g_{2}m_{t}Z_{14}}{\sqrt{2}m_{W}\sin\left(\beta\right)}
\cos\theta_{\tilde t}.
\er
The compositions of $\neut_1$ and mixing of $\st$ are
related with the respective physical states as:
\begin{eqnarray*}
    \neut_1 & = & Z_{11}\tilde{B}+Z_{12}\tilde{W}_{3}+Z_{13}\tilde{H}_{d}+Z_{14}\tilde{H}_{u}\\
    \tilde{t}_{1} & = & \tilde{t}_{L} \cos{\theta_{\tilde{t}}}
    + {\tilde{t}_{R}} \sin{\theta_{\tilde{t}}}
\end{eqnarray*}
where $Z_{ij}$ are the mixing elements in the neutralino sector and
$\theta_{\tilde{t}}$ is the mixing angle between $\tilde{t}_{L}$ 
and $\tilde{t}_{R}$ states.
The couplings,$\gL$ and $\gR$ receive contribution from both the gauge
and Higgs sectors via the composition of neutralino~\cite{Drees:873465}.
As we know, the gauge interaction conserves 
chirality and Yukawa coupling flips it, hence,
the wino ($\tilde{W}_3$) and
bino ($\tilde{B}$) components in $\neut_1$ will preserve the
chirality of interacting fermions
while the Higgsino ($\tilde{H}_u$ and $\tilde{H}_d$) components will flip it.
This scenario is presented in Table~\ref{tab:stopcomp}.
The polarization of the top quark produced in the $\st$ decay, in the rest 
frame of the decaying top squark, can be calculated easily, and 
is given by \cite{Boos:2003vf,Mahlon:1999gz,Schwienhorst:2010je,Perelstein:2008zt,Belanger:2012tm,Shelton:2008nq}:
\begin{eqnarray}
    {\cal P}_{0} & = & \frac{\left(\left|g^{\tilde{t}_{1L}}\right|^{2}-\left|g^{\tilde{t}_{1R}}\right|^{2}\right){{\cal K}^{1/2}}\left(1,\frac{m_{t}^{2}}{m_{\tilde{t}_{1}}^{2}},\frac{m_{\tilde{\chi}_{1}^{0}}^{2}}{m_{\tilde{t}_{1}}^{2}}\right)}{\left(\left|g^{\tilde{t}_{1L}}\right|^{2}+\left|g^{\tilde{t}_{1R}}\right|^{2}\right)\left(1-\frac{m_{t}^{2}}{m_{\tilde{t}_{1}}^{2}}-\frac{m_{\tilde{\chi}_{1}^{0}}^{2}}{m_{\tilde{t}_{1}}^{2}}\right)-4\frac{m_{t}m_{\tilde{\chi}_{1}^{0}}}{m_{\tilde{t}_{1}}^{2}}\text{Re}\left(g^{\tilde{t}_{1L}}g^{\tilde{t}_{1R}*}\right)}
    \label{eq:toppol}
\end{eqnarray}
Where,
\begin{eqnarray*}
    {\cal K}\left(x,y,z\right) & \equiv & {x^{2}+y^{2}+z^{2}-2xy-2yz-2zx}
\end{eqnarray*}
The above expression was derived using the definition:
$$
{\cal P}_0 = \frac{\# t (\lambda_t =1) - \# t(\lambda_t = -1)}{\# t (\lambda_t =1) + \# t(\lambda_t = -1)}
$$
where $\lambda_t$ is the helicity of the top and \# refers to the number of events.

We see that the polarization of the top produced in the $\st$ decay, 
depends on the the  compositions of $\neut_1$ as well as the on $L$-$R$ 
mixing in the $\st$, along with the masses  $m_{\neut_1},   m_{\tilde{t}_{1}}$ 
and $m_t$. The last dependence comes through the kinematic functions, 
and also through the couplings $\gL$ and $\gR$.
\begin{table}
    \begin{center}
        \caption{
            Chirality of top quark from $\st$ decay (\eqnref{eqn:stopdecay})
            for different compositions of neutralino and top squark states.
        }
        \label{tab:stopcomp}
        \begin{tabular}{|c|c|c|}
            \hline
            $\neut_1$ & $\tilde{t}_{1}$     & $t$ chirality\tabularnewline
            \hline 
            Bino like or Wino Like & $\tilde{t}_{L}$ & $ t_{L}$\tabularnewline
            & $\tilde{t}_{R}$ & $ t_{R}$\tabularnewline
            \hline 
            Higgsino like          & $\tilde{t}_{L}$ & $ t_{R}$\tabularnewline
            & $\tilde{t}_{R}$ & $ t_{L}$\tabularnewline
            \hline
        \end{tabular}
    \end{center}
\end{table}
We study the implication of the  kinematic observables
discussed in the last section, namely $z_k$, $Z_b$ and $\cos\theta^*$,
by considering top quarks produced in the
decay of top squark produced in pair in the process,
\begin{equation}
\begin{array}{ccccccc}
pp & \rightarrow &
\tilde{t}_{1}\bar{\tilde{t}}_{1} & \rightarrow &
t\bar{t} {\neut_1} \neut_1 & \rightarrow & \left(bjj\right)\left(b\ell\nu_{\ell}\right)
\neut_1 \neut_1
\end{array} \label{eq:stopeventtype}.
\end{equation}
Clearly the $\st$ and $\bar \st$ are not necessarily produced at rest and hence the polarization of the decay top quarks can be different from that given by \eqnref{eq:toppol}. 
However, the polarization in the frame where the $\st$ is moving can be obtained from it in a very transparent way, if need be,
in terms of that in the rest frame of the decaying top,
as discussed in Ref. \cite{V.:2016wba}

We select the final state containing one isolated lepton
(either electron or muon) and at least one b jet with large
amount of missing energy due to the presence of weakly interacting
neutralinos.
Leptons are considered in order to have less contamination
from the SM backgrounds, mainly from QCD production,
and, of course to focus only on hadronic decay mode of
the other top quark avoiding recombinatorial issues
\cite{Plehn:2010st}.
We set relevant parameters as,
\[
\begin{array}{ccccccccccc}
m_{\neut_1} & = & 100\text{ GeV} & ;
& m_{\tilde{t}_{1}} & = & 1\text{ TeV} & ;
& \tan \beta & = & 10
\end{array}
\]
Events are generated using {\mg}
with the minimal supersymmetric standard model(MSSM),
appropriately setting
the composition of $\tilde{t}_1$ and $\neut_1$.
The complete chain described in
\eqnref{eq:stopeventtype} is generated
using the matrix element to ensure
accurate preservation of spin effects.
In simulation, the events are selected as:
\begin{itemize}

    \item
    At least 1 hard ($p_T > 20$ GeV) and isolated lepton
    with $|\eta|<2.5$, isolation is ensured by demanding
    the momentum fraction
    $\frac{\sum_{\left(\Delta R<0.3\right),i}p_{T_{i}}}
    {p_{T_{\ell}}}<0.3$.

    \item
    Hard missing transverse momentum with
    $p_{T}^{\text{miss}}>30$ GeV.

    \item
    Top jet is tagged by clustering the particles into
    C/A \cite{Dokshitzer:1997in} fatjets of $R=1.0$ using {\fastjet}
    and then passed through {\toptag} and demand the event with at
    least one top tagged fat jet.
    If there are more than one top tagged fatjet,
    we use the top jet which is best
    reconstructed i.e by checking the reconstructed mass 
    closer to the physical mass of the top quark.

\end{itemize}

The impact of top polarization on the energy fraction variables,
{$Z_b$(\eqnref{eq:Zk})} and {$z_k$(\eqnref{eq:kronzk})},
is presented in \figref{fig:StopZb}
and \figref{fig:Stopzk}.
Here, we assume $\neut_1$ to be pure bino. 
The  expected top polarization  in the rest frame of $\st$, for the cases 
where $\st$ is pure $\tilde t_L$ or $\tilde t_R$,  for the choice of masses of the $\st$ and the $\neut_1$, is $\mp 0.9994$ respectively. Due to the rather 
large mass of the $\st$ considered, the polarization in the laboratory frame where $\st$  is produced  is not very different from this value as the $\st$ is produced mostly at rest \cite{V.:2016wba}.

\begin{figure}[H]
    \begin{center}
        \includegraphics[width=0.5\textwidth]
        {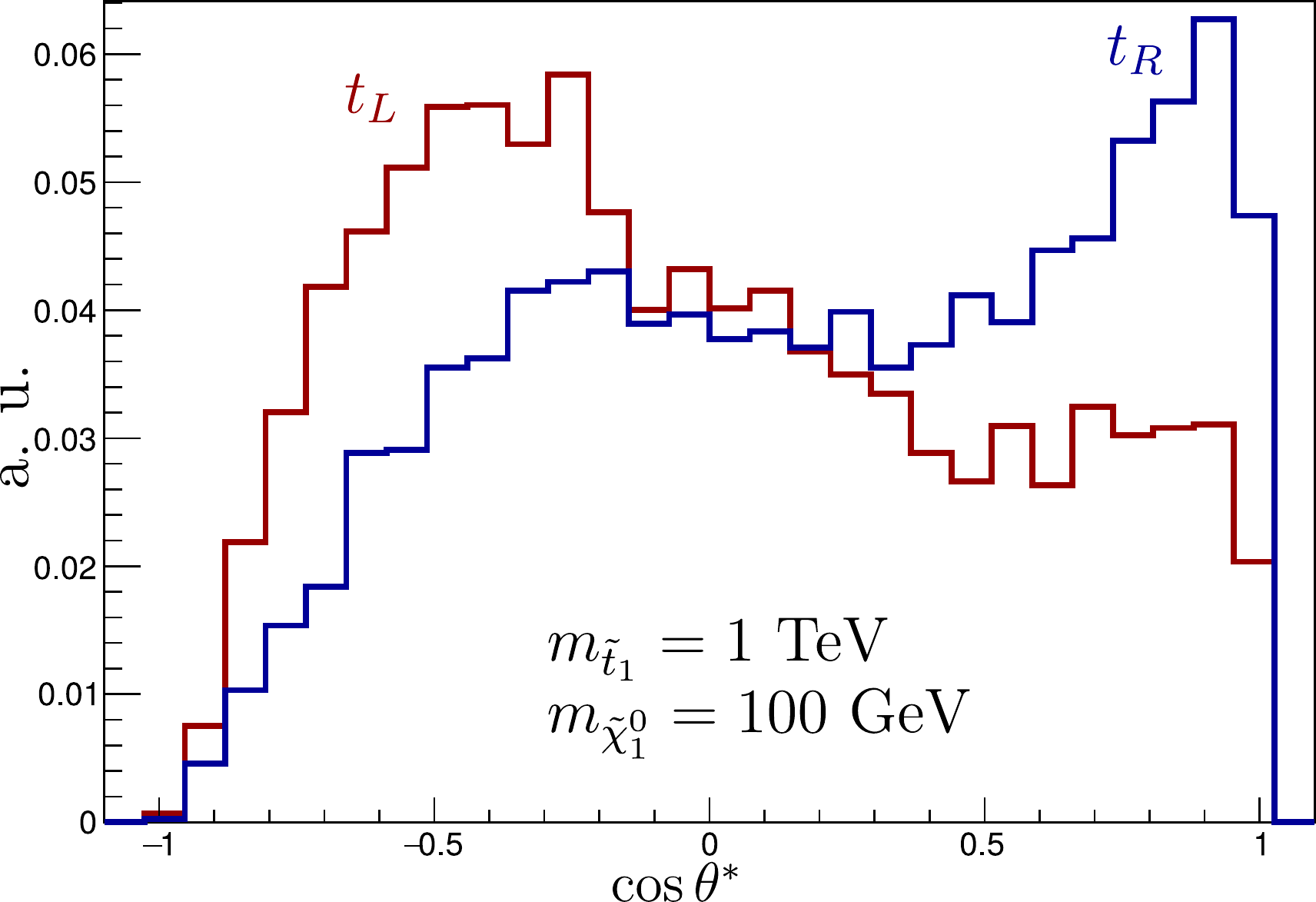}
        \caption{
            The distribution of $\cos\theta^*$
            (\eqnref{eq:ourobs}) corresponding to the case:
            LSP is pure bino ($\neut_1=\tilde{B}$)
            and $\tilde{t}_1$ is either pure $\tilde{t}_L$ (red)
            or $\tilde{t}_R$ (blue).
        }
        \label{costhetaststop}
    \end{center}
\end{figure}

\begin{figure}[t]
    \begin{center}
        \includegraphics[width=0.49\textwidth]
        {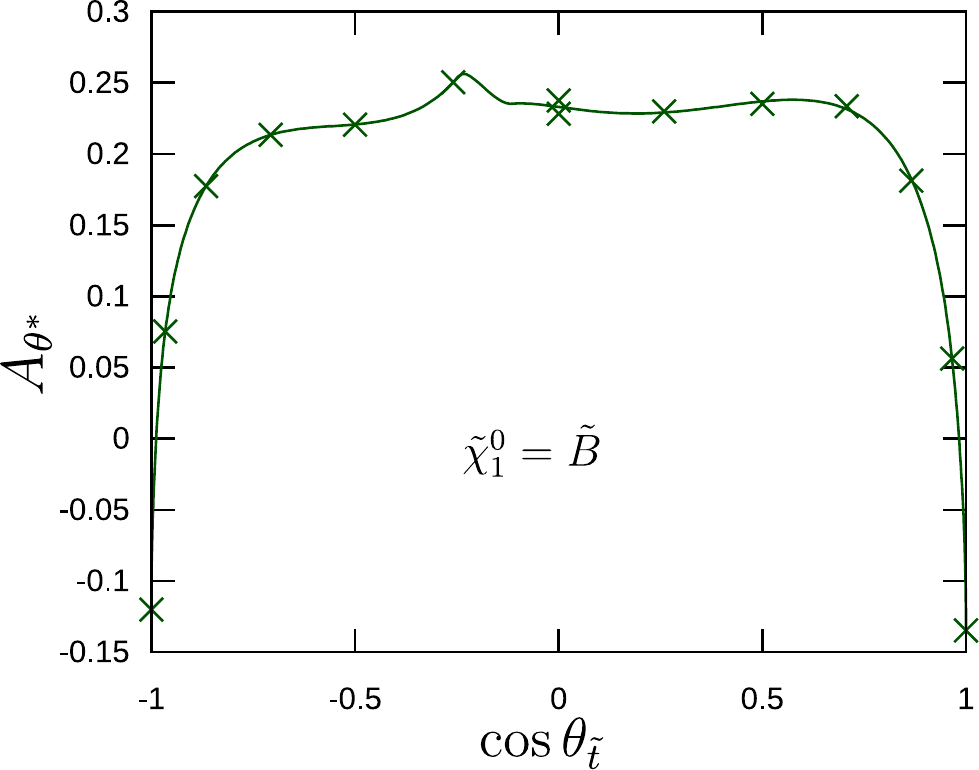}
        \includegraphics[width=0.49\textwidth]
        {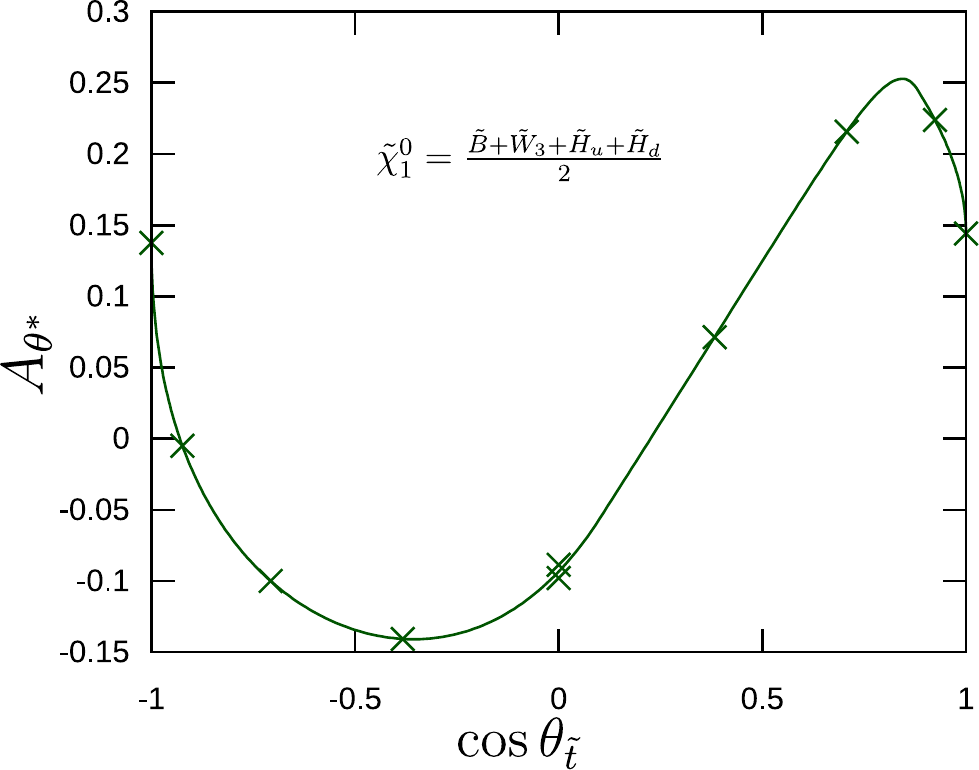}
        \caption{
            Variation of asymmetry with
            the composition of $\tilde{t}_1$ for the cases of pure bino 
like(left) and
            equally mixed $\neut_1$ states(right).
        }
        \label{fig:varythetastop}
    \end{center}
\end{figure}

Clearly the distributions show different pattern for
left and right handed top quarks coming from top squark decay 
in the two cases.
As expected, events are crowded towards higher(lower)
region for left(right) handed top quarks.
Perhaps, a selection as $z_k (Z_b) \ge 0.5$ can clearly distinguish the region dominated by left handed top quarks.
Hence, measurement of these variables clearly indicate
the state of polarization of top in this decay channel.
As presented before, more robust variable in this context
is $\cos\theta^\star$, shown in
\figref{costhetaststop}.
Evidently, this distribution demonstrate as before, 
a very clear distinction between left and right polarized 
top quark. 
Based on this angular variable, 
we calculate the asymmetry as defined by
\eqnref{eq:asym} \cite{Belanger:2012tm}.
It is obvious that this asymmetry is expected to be very sensitive to 
the composition of neutralino and the chirality of
$\st$(i.e. $\theta_{\tilde t}$).
We systematically study the variation of $A_{\theta^*}$.
The dependence of it on $\theta_{\tilde t}$
is studied fixing the composition of $\neut_1$ and
results are presented in \figref{fig:varythetastop} for a pure bino like
($Z_{11}=1$) LSP scenario(left) and for equal contents
($Z_{11}=Z_{12}=Z_{13}=Z_{14}=\frac{1}{2}$)
of four states in $\neut_1$(right).

\begin{minipage}{0.48\textwidth}
    \begin{figure}[H]
        \includegraphics[width=1.0\textwidth]
        {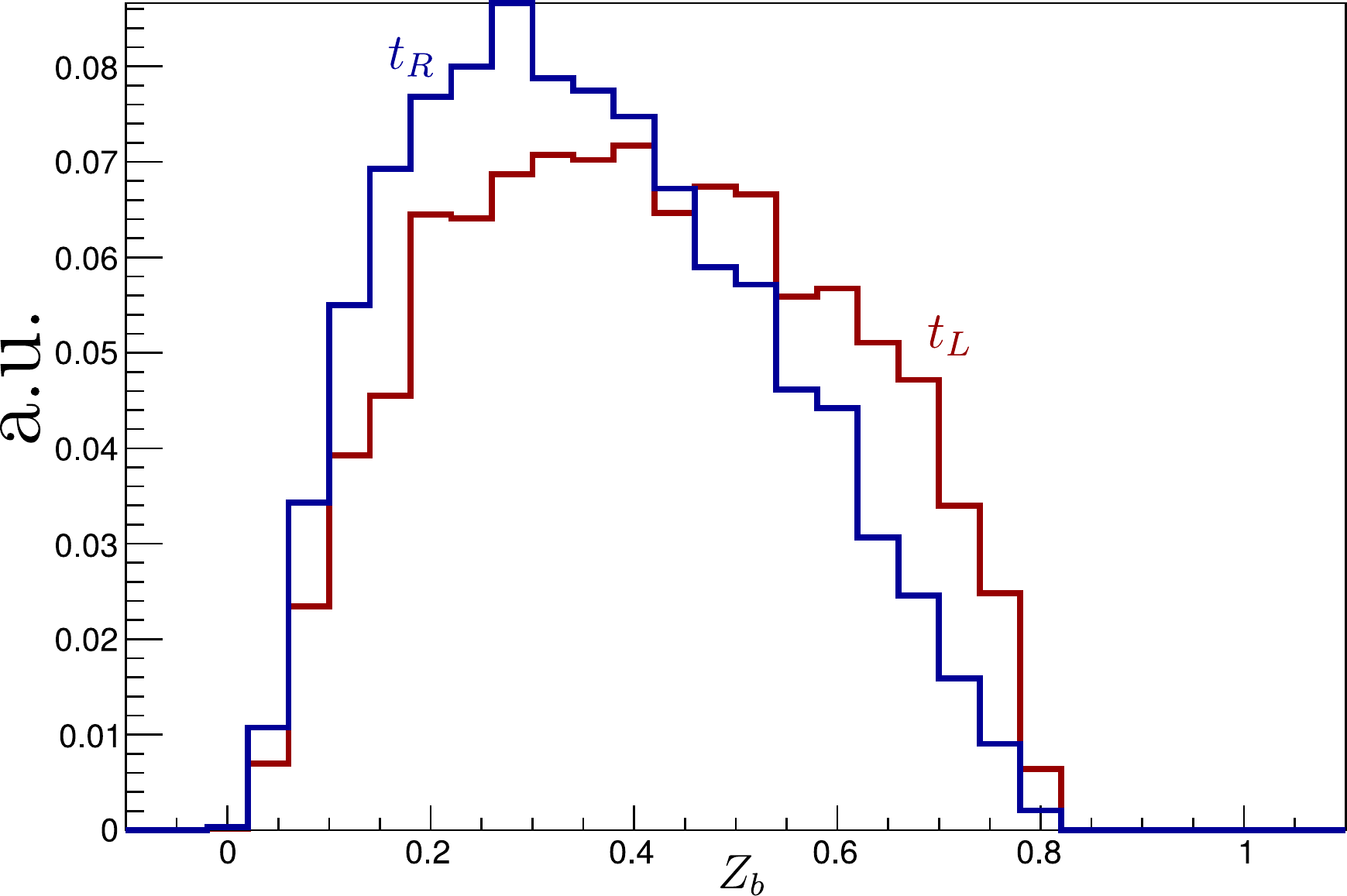}
        \caption{
            The $b$ quark energy fraction $Z_b$
            for the event presented by \eqnref{eq:stopeventtype},
            LSP is pure bino ($\neut_1=\tilde{B}$) and $\tilde{t}_1$ is either
            pure $\tilde{t}_L$ (red) or $\tilde{t}_R$ (blue).
        }
        \label{fig:StopZb}
    \end{figure}
\end{minipage} \hspace{0.02\textwidth}
\begin{minipage}{0.48\textwidth}
    \begin{figure}[H]
        \includegraphics[width=1.0\textwidth]	
        {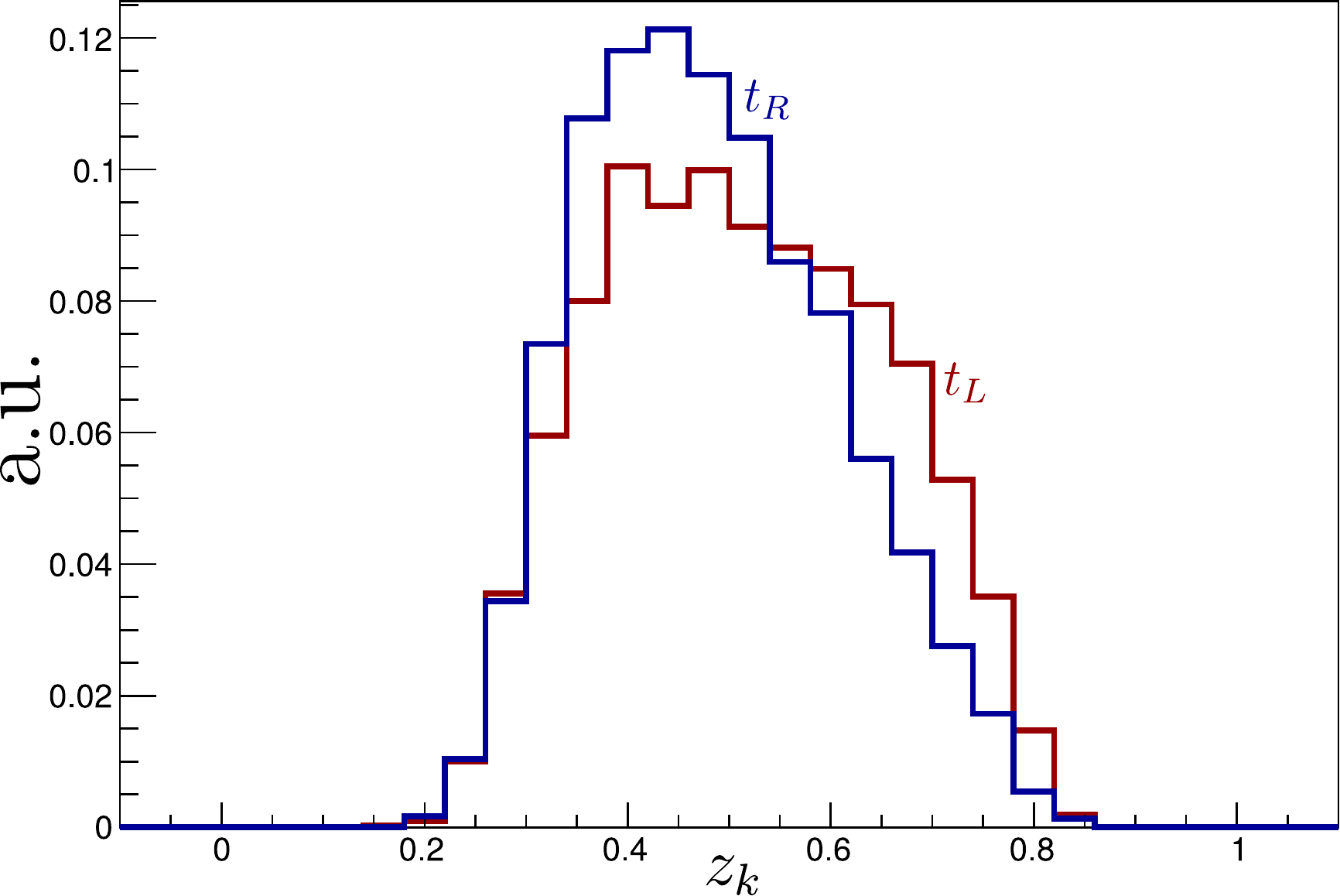}
        \caption{
            Same as \figref{fig:StopZb} but for $z_k$.
        }
        \label{fig:Stopzk}
    \end{figure}
\end{minipage}\\

\begin{figure}[t]
    \begin{center}
        \includegraphics[width=0.49\textwidth]
        {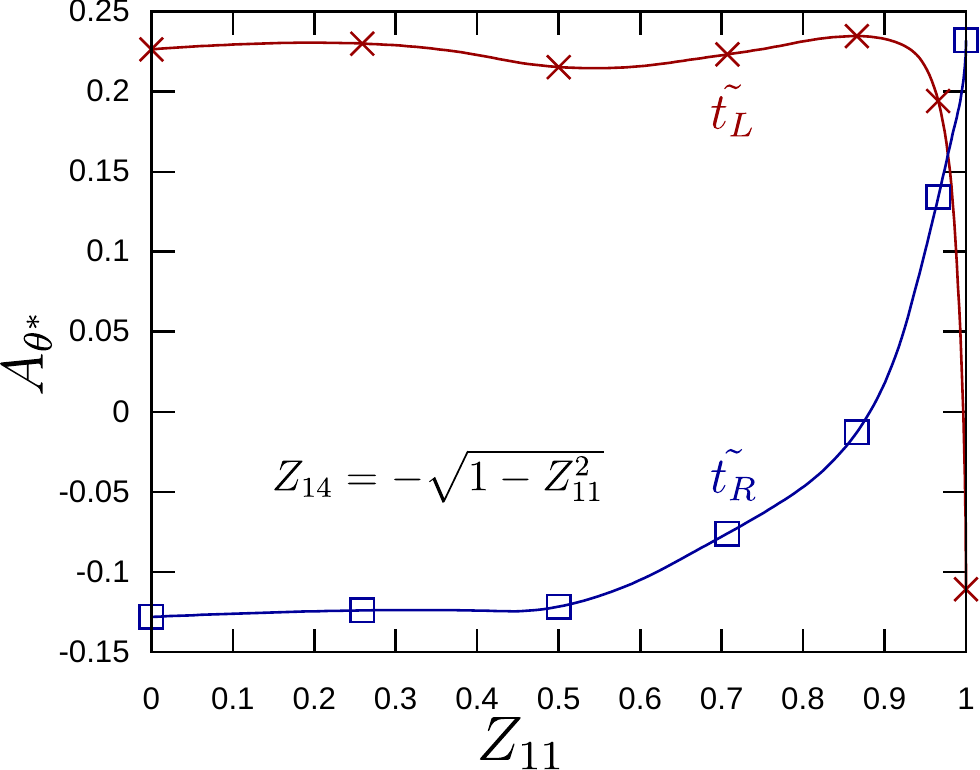}
        \includegraphics[width=0.49\textwidth]
        {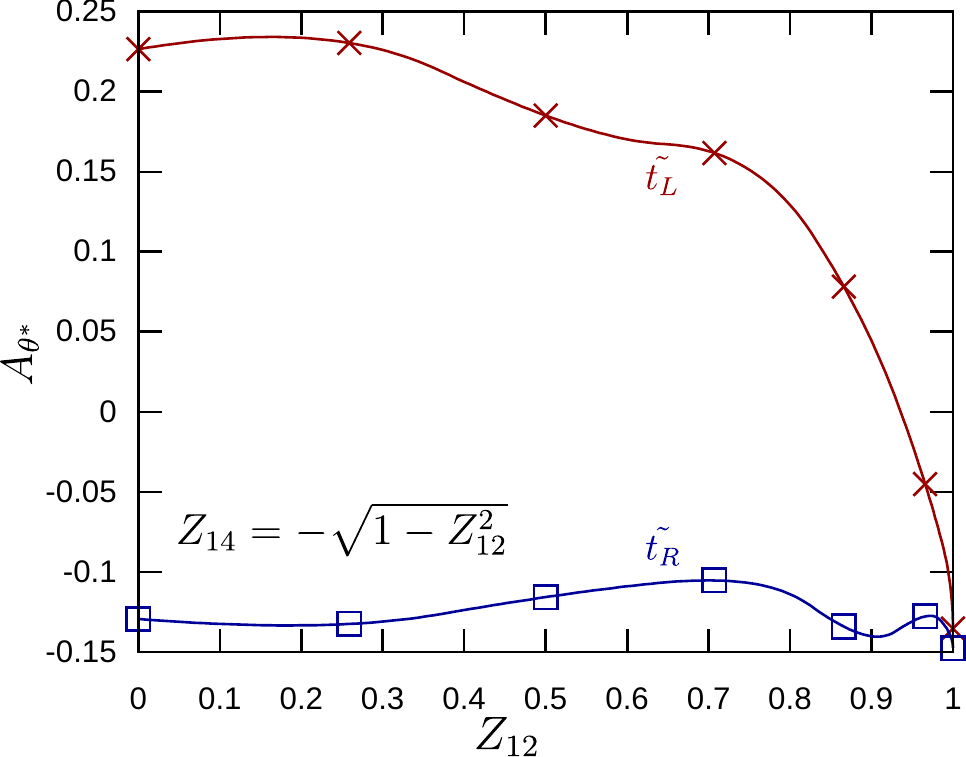}
        \caption{
            Variation of
            asymmetry (\eqnref{eq:asym}) with the
            composition of bino and wino like $\neut_1$ for the cases of
            $\tilde{t}_{1}=\tilde{t}_{L}$ and $\tilde{t}_{1}=\tilde{t}_{R}$.
        }
        \label{fig:Varchi10}
    \end{center}
\end{figure}

It is worth to mention here some features of the distributions shown 
in this \figref{fig:varythetastop}. 
For a pure bino like case (left plot), and 
larger values of $\left| \ctht \right| \sim 1 $, $\st$ and $\neut_1$ 
couplings become, $g^{\tilde{t}_{L}} \gg  g^{\tilde{t}_{R}}$, i.e produced top 
quark is pre-dominantly left handed, leading to $\cths$ 
negative, (see Fig.~\ref{costhetaststop}), and hence making $A_{\theta^*}$ 
negative(Eq.~\ref{eq:asym}). With the decrease of magnitude of
$\ctht$, the right handed coupling part $g^{\st}_R$ gradually becomes
important, leading to the production of both left and right handed top quarks.
Since, the coupling of $\tilde{t}_R$ with $\neut_1$ is twice that 
of $\tilde{t}_L$ due to the hyper charge, the abundance of
right handed top is more than left handed top quarks. 
Hence, $A_{\theta^\star}$ turns out to be positive, and does not 
change much even with the change of $\ctht$, and finally again becomes negative
for $\ctht \sim 1$. 
In the case of equal mixtures of
gauginos and Higgsinos the variation of $A_{\theta^\star}$ is shown
in \figref{fig:varythetastop}(right). Again for larger values of 
$|\ctht|$ (i.e, $\st \sim \tilde t_L$), 
the Higgsino contribution to the coupling($g^{\st}_R$) becomes
dominant due to the dependence on the top quark mass, and since the
Yukawa couplings flip the chirality, the top quark tends to be  
right handed which makes $A_{\theta^\star}$ positive, as it is clearly observed.
For intermediate values of $\ctht$, the left handed top quarks are also
produced making the asymmetry negative, and for $\ctht>0$ and beyond, 
again the population of $t_R$ goes up, making asymmetry positive.
In Fig.~\ref{fig:Varchi10}, the variation of asymmetry 
on the composition of neutralinos for a given chirality of 
$\st$ is shown. In the left plot, the magnitude of $\tilde B$ content in
$\neut_1$ is varied for both cases of $\tilde t_L \left( \ctht=1 \right)$ and 
$\tilde t_R \left( \ctht=0 \right)$ production assuming $Z_{12}, Z_{13}=0$ and 
$Z_{14} = -\sqrt{1-Z_{11}^2}$.
Due to much larger Higgsino coupling, the 
interaction between top squark and Higgsino like LSP dominates 
as compared to the gaugino case. Hence the top quark from the decay 
of left(right) handed like ($\st$), 
becomes right(left) handed.
It implies that for $\cos \theta_{\tilde{t}} =1(0)$, the 
asymmetry expected to be positive(negative). With the
increase of $Z_{11}$, the Higgsino coupling becomes less important,
and hence left(right) handed like $\st$ preferably decays to left(right) 
handed top quark resulting in a gradual flipping of the sign of asymmetry 
as shown in the left figure. In the case of variation of asymmetry  
between the cases 
of bino and Higgsino like LSP, the steeper curve for $\tilde{t}_L$ compared to
$\tilde{t}_R$ is again a consequence of the higher hypercharge of
$\tilde{t}_R$.
In \figref{fig:Varchi10}(Right) the variation of $A_{\theta^*}$ with 
wino($Z_{12}$) and Higgsino like LSP($Z_{14}$) setting $Z_{11}=Z_{13}=0$
are presented.  
For lower range of
$Z_{12}$, the main contribution
comes due to the Higgsino like coupling, and it goes down with the 
increase of $Z_{12}$.
Consequently, in this case the asymmetry for $\tilde t_L$
changes faster in comparison to
$Z_{11}$ variation(left),
which can be attributed to the fact that the
iso-spin interaction having larger
magnitude than the hyper-charge interaction.
It is obvious that the variation  
of asymmetry for $\tilde{t}_R$ case is
essentially unaffected, since it does not couple to the winos.
Obviously, if the composition of
$\neut_1$ be mixed
then the variation of $A_{\theta^*}$
is expected to
be in between the
two extreme cases as shown in \figref{fig:Varchi10}.
Undoubtedly, the measurement of asymmetry is found to be a robust tool
to probe the polarization of top quark,
and the nature of couplings involved in the process. 
In future, if top squark is discovered, and its polarization is
measured, then it will allow us
to constrain the compositions of top squark with respect to the mixings
of the neutralino, provided we have some idea about the neutralino 
sector as well.

\section{Top polarization in RS models}
\label{sec:RSmodel}
In this section we study the production of polarized top quark in extended
Randall-Sundrum(RS) models~\cite{Aquino:2006vp,Geng:2018hpq}, where top quark can be produced
from the decay of the heavy resonance of mass of few TeV.
These resonances can be Kaluza Klein (KK)-excited states of SM particles.
Various KK-excited states can be present in different RS models depending on how the SM-states propagate in the bulk~\cite{Gherghetta:2010cj}.
Moreover, the couplings of KK-states to the SM fermions 
may not be universal. These extended RS models have an advantage over the original RS model~\cite{Randall:1999ee} of having a natural solution of 
Yukawa hierarchy problem \cite{Pomarol:1999ad,Gherghetta:2000qt,Grossman:1999ra}. We emphasize here on the fact that,  
one can have $t \bar{t}$ or single top as the final state depending on which KK state is the mediator.
To represent these two classes, we 
consider  KK-gluon (flavor violating ) and KK-graviton ($G_{KK}$) decay
where we get single top and $t \bar t$ final state respectively. We also consider the case when KK-gluon state decays into a top-antitop pair for comparison. 

\begin{figure}[t]
    \begin{center}
        \includegraphics[width=0.49\textwidth]{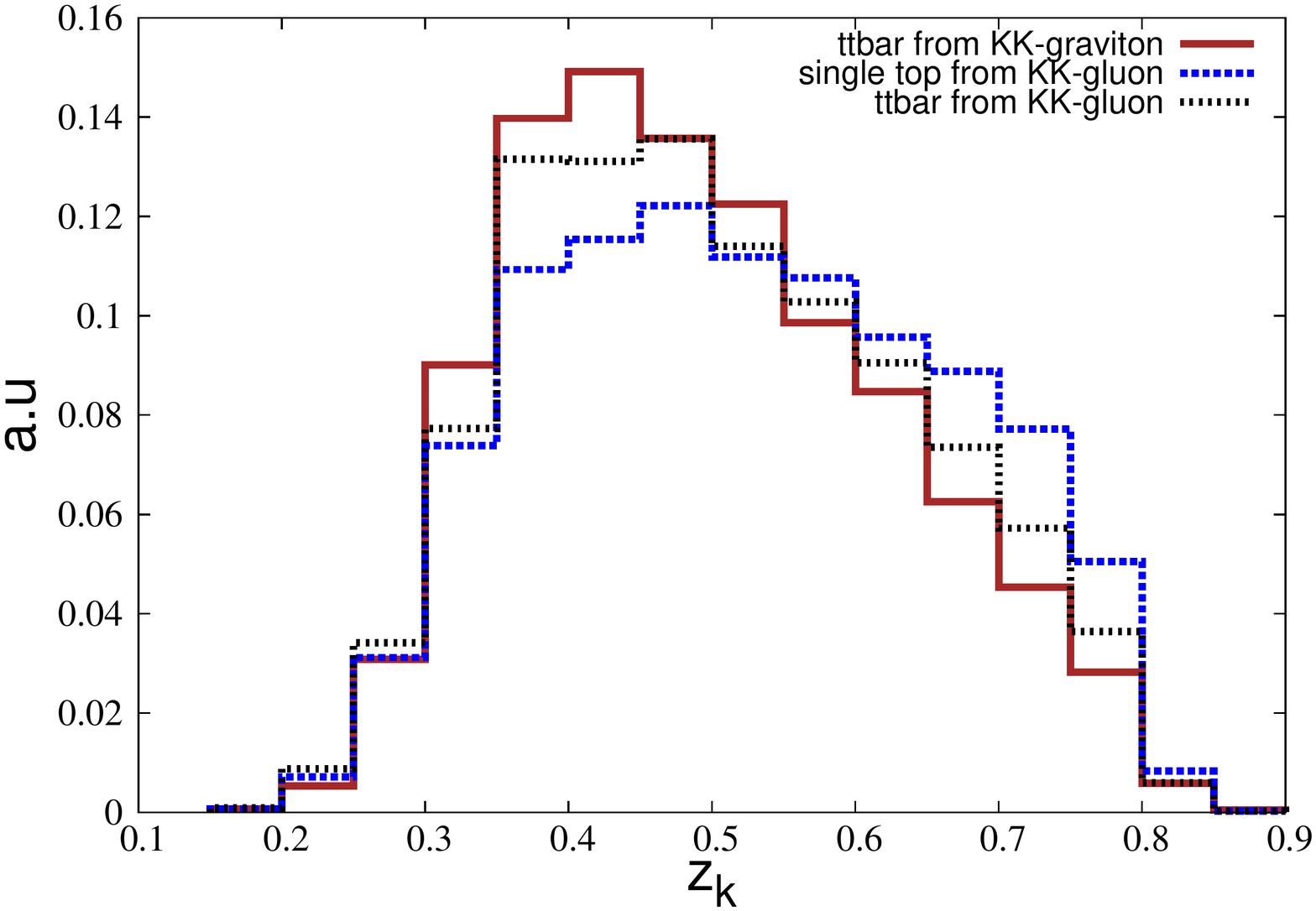}
        \includegraphics[width=0.49\textwidth]{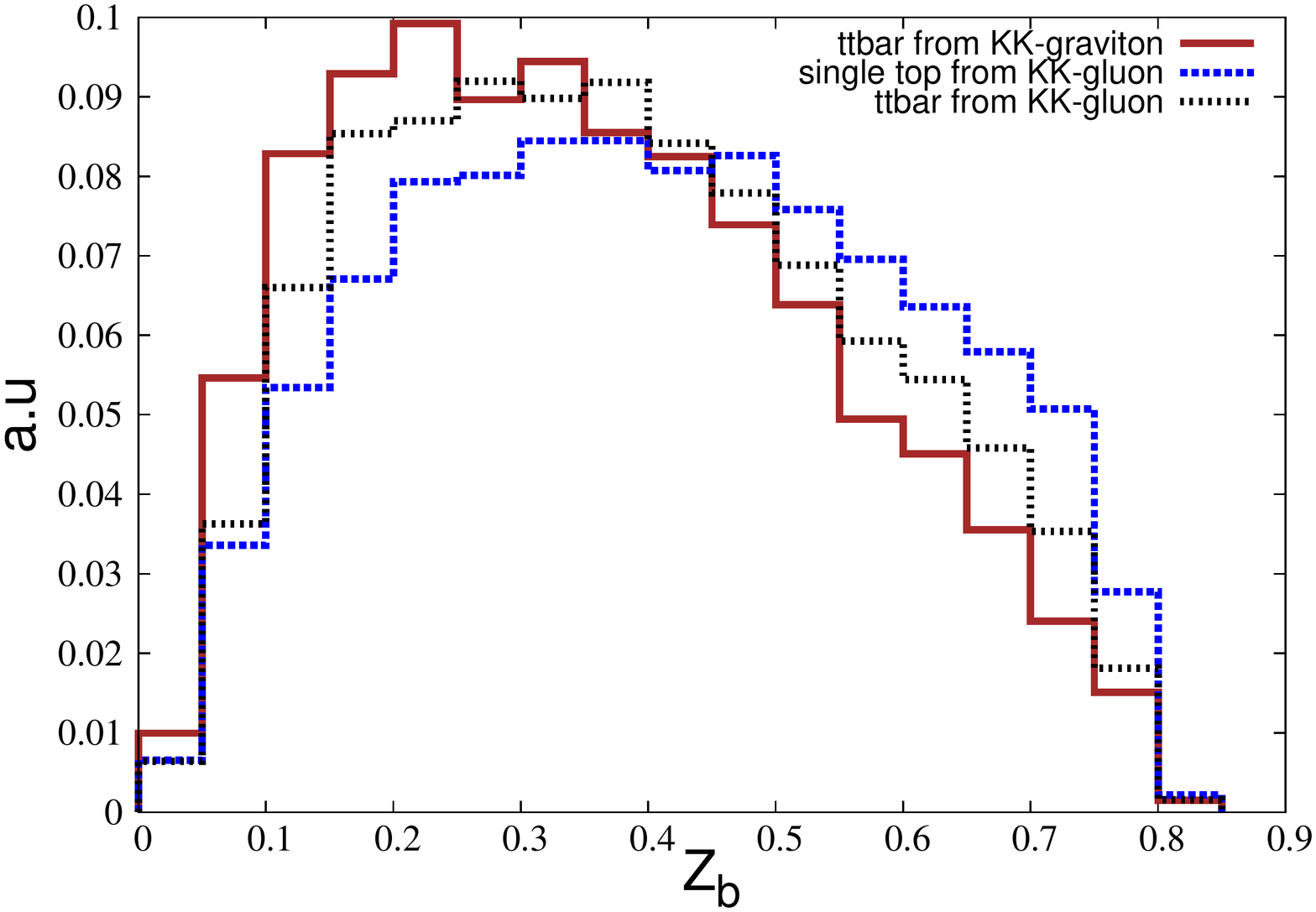} \\
        \includegraphics[width=0.49\textwidth]{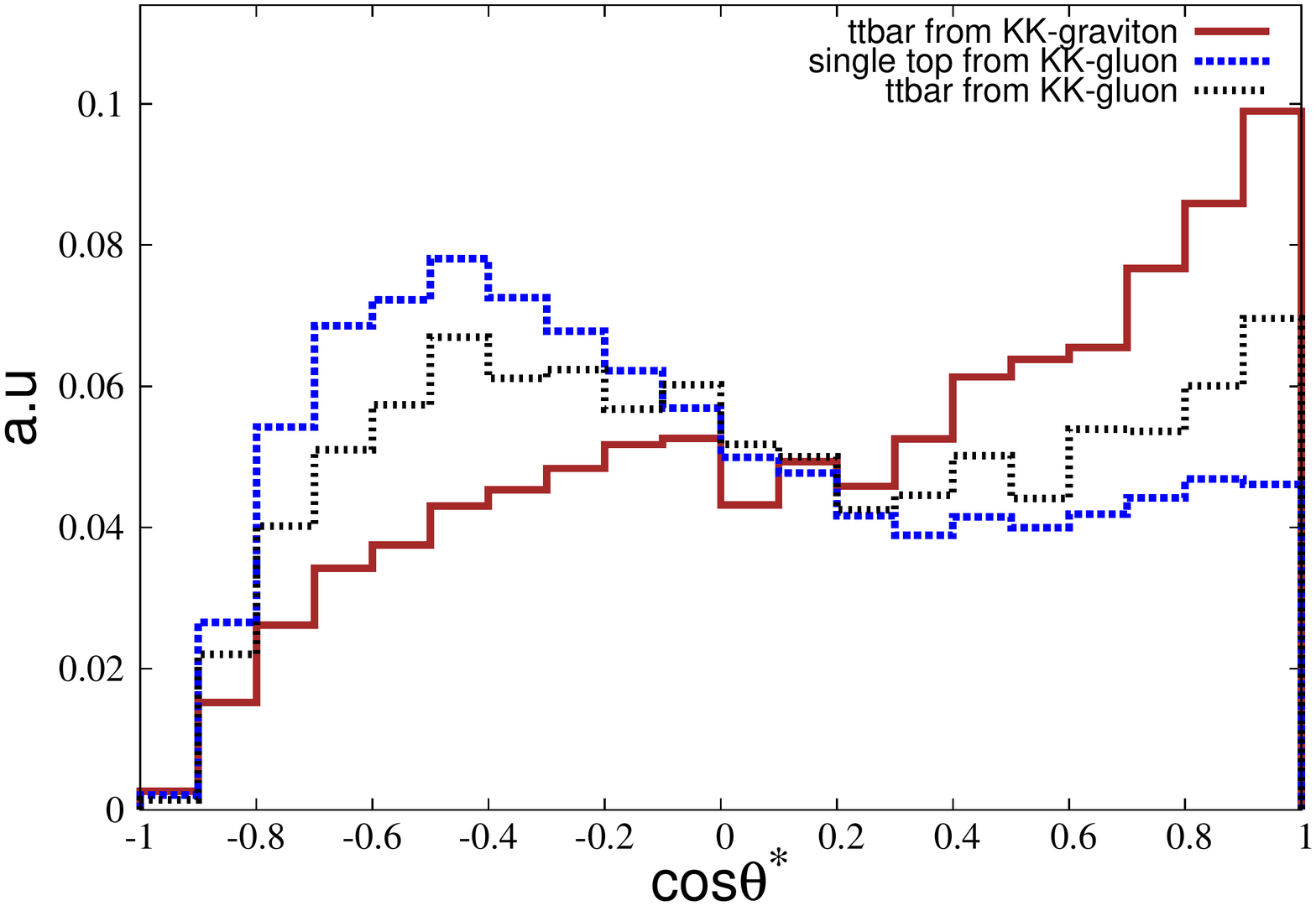}
        \caption{
            The $z_k$(top left), $Z_b$(top right) and $\cos \theta^*$(bottom center) distribution of reconstructed top jets for single top and $t\bar t$ pair production from KK-gluon of mass 4 TeV and $t \bar t$ pair from KK-graviton of mass 4 TeV.
        }
        \label{RSkkgfigs}
    \end{center}
\end{figure}

First we consider, a single top production, in
``Kaluza-Klein Gluon Model'' \cite{Aquino:2006vp}.
This model allows flavor-violating neutral current interactions of the KK gluon.
The flavor-violating localization of fermions induces flavor-violating interactions of the KK-gluon. Therefore in this model the KK-gluon will have flavor-violating neutral coupling of the form $\bar{t} \gamma_{\mu} q g^{*{\mu}}$ (where $q$ denotes light quarks), along with the flavor-conserving neutral coupling $\bar{t} \gamma_{\mu} t g^{*{\mu}}$ ~\cite{Aquino:2006vp}.  
We have considered a specific case when only
left-chiral couplings are allowed in case of single top production and in particular focus
on the flavor-violating decay of the KK-gluon to a top quark and a charm quark.
We have analyzed both $t \bar t$ and single top production in this model. We generate following  processes,
\begin{eqnarray*}
    pp \rightarrow g_{KK} &\rightarrow& t_L \bar{c}_L \rightarrow \text{Hadronic final states}~~~~(\text{both s and t-channel)}\\
    pp \rightarrow g_{KK} &\rightarrow& t \bar t \rightarrow \text{Semi-leptonic final states}~~~~(\text{s-channel)}
\end{eqnarray*}
We used the available {\tt FeynRules} model\cite{Drueke:2014pla,Chivukula:2014pma}
file and events are generated with {\mg}.
Showering and hadronization are done using {\tt Pythia6}\cite{Sjostrand:2006za}.
For single top events, hadronic final state is considered while for $t\bar{t}$ events,
the following criteria is used to select events:
\begin{itemize}
    \item At least one hard ($p_T > 20$ GeV) and isolated lepton is demanded. The isolation is ensured by demanding $\frac{\sum_{i} p_T^i(\Delta R < 0.3)}{p_T^l} < 0.3$.
    \item A cut on the missing transverse energy has also been applied i.e. $\cancel{E_T} > 30$ GeV. 
\end{itemize}
We passed the selected events through {\toptag} to tag the top
jets out of fatjets constructed with $R=1.0$.
The $t \bar t$ final state produced via s-channel mediation of KK-gluon
are unpolarized and single top produced via s- and t-channel exchange 
of KK-gluon is left-chiral.
In \figref{RSkkgfigs}, we present the $z_k$, $Z_b$ and $\cos \theta^*$ distribution
for $t \bar t$ and single top .
We have considered KK-gluon mass of 4 TeV which is allowed by the current
experimental bound~\cite{Sirunyan:2018ryr}. 

We also consider $t \bar t$ production from KK-graviton in top-philic model 
discussed in Ref. \cite{Geng:2018hpq}. In this model, right-chiral
top quark will be localized near the infra-red(IR) brane and the 
KK-graviton is also localized near the IR brane.
Therefore KK-graviton will have dominant coupling to
right-chiral fermions and hence the produced top quark will
be right-chiral. The corresponding Lagrangian is as follows:
\begin{equation}
{\cal L}_F= -\frac{1}{\Lambda} G^{\mu\nu}{\cal T}^F_{\mu\nu}, \nonumber
\end{equation}
where $G^{\mu\nu}$ is the graviton field, ${\cal T}^F_{\mu\nu}$
is the energy-momentum tensor of the fermion fields.
\begin{eqnarray}
\label{LagB1}
{\cal T}^F_{\mu\nu} &=& \sum_{f=u,d,l}\left[\frac{i}{4}\bar{f}_R (\gamma^\mu D^\nu + \gamma^\nu D^\mu) f_R -\frac{i}{4} (D^\mu \bar{f}_R \gamma^\nu + D^\nu \bar{f}_R \gamma^\mu) f_R \right.\nonumber\\
&& \left.
- i\eta^{\mu\nu}\left\{\bar{f}_R \gamma^\rho D_\rho f_R -\frac{1}{2}D^\rho(\bar{f}_R \gamma_\rho f_R)\right\} \right]\,,
\end{eqnarray}    
here $\eta_{\mu\nu} = \mathrm{diag}(1, -1, -1, -1)$ is the
Minkowski metric tensor, and
$D_\mu = \partial_\mu + i (2/3) g_1 B_\mu + i g_s G^a_\mu$
is the covariant derivative corresponding to each fermion. We consider the process
$p p \rightarrow G_{KK} \rightarrow t \bar t, (t \rightarrow W^+ b, W^+ \rightarrow j j),
(\bar t \rightarrow W^- \bar b, W^- \rightarrow l^- \bar{v_l})$ + h.c.
In this case also the event generation process and selection criteria are same as the KK-gluon case.

In \figref{RSkkgfigs}, we also show the distributions for $z_k$, $Z_b$ 
and $\cos\theta^*$
for KK-graviton mass of 4 TeV. This value of KK-graviton mass is consistent 
with the experimental results \cite{ATLAS:2018tpf}.
In this model, the KK-graviton couples only to the right-chiral top quark 
as mentioned earlier. Therefore, the tops produced through $t \bar t$ pair 
production from the KK-graviton decay, will be right-chiral. 
A comparison among the three curves in each panel of \figref{RSkkgfigs}, 
indicates that the discriminating potential of $z_k$ variable is   
less promising than the other two variables considered. The $b$-subjet 
energy fraction, $Z_b$ shows a better
discriminating performance between left-chiral, right-chiral and unpolarized tops which are produced from different resonances. It is clear from this figure that the best discriminator and a robust observable in the tagger environment is indeed $\cos \theta^*$. In case of left-chiral top produced from KK-graviton, 
the distribution is more populated in the lower range of $\cos \theta^*$ and 
for right-chiral top quark produced from KK-gluon it is peaked around higher 
range of $\cos \theta^*$. For unpolarized top from KK-gluon the distribution is comparatively more uniformly distributed across the whole range of $\cos \theta^*$. It is clear from these plots that the variable $\cos \theta^*$ not only gives us information about the chiral structure of the top quark coupling, but also helps us to distinguish between different models. The KK-graviton and KK-gluon have different coupling structures with the top quark. But as they produce top quark with different polarization, these observables can also be utilized to distinguish between these two types of resonances.

Next we extend this study to top-philic model and
consider non-zero coupling of KK-graviton with both
left- and right-chiral top quark,
and
calculated the
asymmetry for the $\theta^*$ observable,
as we did for top production from $W'$ decay and stop decay in MSSM.
We have introduced parameters $k_L$ and $k_R$ in the Lagrangian, as defined,
\begin{equation}
{\cal L}_F= -\frac{1}{\Lambda} G^{\mu\nu}{\cal T}^F_{\mu\nu}, \nonumber
\end{equation}
with
\begin{eqnarray}
\label{LagB2}
{\cal T}^F_{\mu\nu} &=& \sum_{f=u,d,l,\nu_l}\left( \frac{k_R}{\sqrt{k_L^2+k_R^2}}\left[\frac{i}{4}\bar{f}_R (\gamma^\mu D^\nu + \gamma^\nu D^\mu) f_R -\frac{i}{4} (D^\mu \bar{f}_R \gamma^\nu + D^\nu \bar{f}_R \gamma^\mu) f_R \right. \right.\nonumber\\
&& \left.
- i\eta^{\mu\nu}[\bar{f}_R \gamma^\rho D_\rho f_R -\frac{1}{2}D^\rho(\bar{f}_R \gamma_\rho f_R)] \right]+ \frac{k_L}{\sqrt{k_L^2+k_R^2}}\left[\frac{i}{4}\bar{f}_L (\gamma^\mu D^\nu + \gamma^\nu D^\mu) f_L\right.\nonumber\\&&   -\frac{i}{4} (D^\mu \bar{f}_L \gamma^\nu + D^\nu \bar{f}_L \gamma^\mu) f_L  
- i\eta^{\mu\nu}[\bar{f}_L \gamma^\rho D_\rho f_L -\frac{1}{2}D^\rho(\bar{f}_L \gamma_\rho f_L)] \nonumber\\&&
-\eta_{\mu \nu}
\left(\frac{g_W}{\sqrt 2}V_{ij}{\bar f}_{u_i} \gamma^{\rho} P_L f_{d_j} W^+_{\rho} + \frac{g_W}{\sqrt 2}U_{ij}{\bar f}_{l_i} \gamma^{\rho} P_L f_{\nu_j} W^-_{\rho} + h.c\right)
\nonumber\\
&&  \left. \left.
-  \left(\frac{g_W}{\sqrt 2}V_{ij}{\bar f}_{u_i} \gamma_{\mu} P_L f_{d_j} W^+_{\nu}+ \frac{g_W}{\sqrt 2}U_{ij}{\bar f}_{l_i} \gamma_{\mu} P_L f_{\nu_j} W^-_{\nu} + h.c + (\mu \leftrightarrow \nu)\right)
\right]  \right).\
\end{eqnarray}    
This equation
suggests a new parameterization, in terms of angular variable $\theta_{RS}$ for the calculation of asymmetry, where $\cos \theta_{RS} = \frac{k_L}{\sqrt{k_L^2+k_R^2}}$. When $|k_L| = 1 $ and $k_R = 0$,
implies
$|\cos \theta_{RS}| = 1$, i.e only the left-chiral top couples to the KK-graviton
and while $k_L = 0 $ and $|k_R| = 1$
means
$\cos \theta_{RS} = 0$ so the coupling of right-chiral top to the KK-graviton is non-zero. 
\begin{figure}[H]
    \centering
    \includegraphics[scale=0.35]{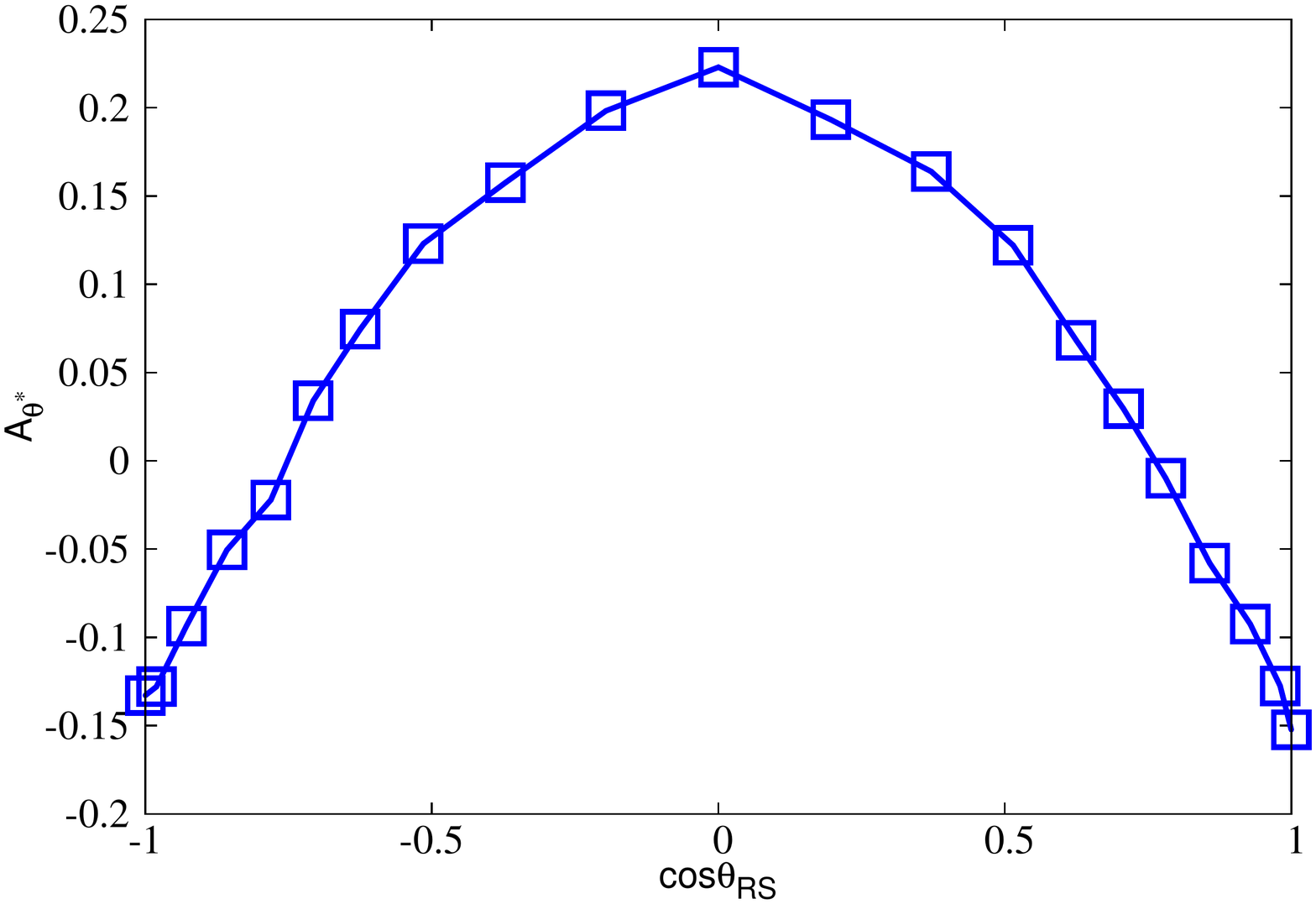}
    \caption{The $A_{\theta^*}$ asymmetry for tagged tops for KK-graviton mass of 3 TeV.}
    \label{asymrs}
\end{figure}
The asymmetry $A_{\theta^*}$ for the tagged tops as a
function of $\cos \theta_{RS}$ is presented in \figref{asymrs}.
The asymmetry is negative for the left-chiral case
and positive for the right-chiral case. In this figure we notice
an interesting feature that the magnitude of asymmetry
$|A(\theta^*)|$ is larger for the right-chiral top than in the
left-chiral case. The reason behind this
is the sharp falling of the $\cos \theta^*$ distribution in
case of left-chiral top quark, when $\cos \theta^* \sim 1$(see Fig~\ref{RSkkgfigs}).
Clearly, the asymmetry is a robust and efficient
probe of new physics which produces polarized top quarks.
Its deviation from the SM prediction will indicate the
presence of polarization-sensitive new physics in the top sector. 
The angular observable
$\cos \theta^*$ and the asymmetry $A_{\theta^*}$ together will
not only help us to look for the new physics but also guide us in
identifying its nature.

\section{Summary}
\label{sec:summary}
Top quark polarization carries the information which can provide deeper 
insight about its couplings with fermions and gauge bosons. 
Hence, probing the coupling of top quark through the measurement of its
polarization, possibly, can indicate the presence of new physics.
The kinematic distributions of decay products from top quark, including angular 
correlations among themselves, are strongly correlated with it's
helicity.
From the conservation of momentum and spin in the hadronic
mode of top decay, it can be shown that the anti-fermion from W decay is
very sensitive to polarization in comparison to other decay products.
However, it is very challenging
to identify this polarization sensitive decay product
for highly boosted top quark of which decay products 
are not very well separated, and come out within a fat jet.  
In this study, we attempt to develop a strategy to study the polarization 
of boosted
top quark focusing on its hadronic decay mode considering a benchmark 
process, \eqnref{eq:wprime}.
The boosted top quark is
tagged using jet substructure method with clear identification
of sub-jets. We identify the subjet corresponding to the 
decay(anti-fermion) product from W decay, and obtain its energy fraction
in the laboratory frame, which is found to be very useful
sensitive variable to distinguish the left and right 
handed top quark. Moreover, availability of techniques to 
tag b jets with very high transverse momentum inside a tagged top jet,
also motivates us to use tagged b-jets as another
object to identify top quark polarization by measuring
its energy fraction in the lab frame. It is to be noted that previously  
b-tagged jets were not used in this context due to the lack of proper 
b tagging tool. It is observed that for hard(soft) 
b-tagged jet or untagged jet, the probability of 
mother top quark to be left(right) handed is more, i,e
hardness of jets clearly distinguishes the left and right handed top quark,
see \figref{fig:BEFrac} and \ref{fig:cpsthetastar}.
Furthermore, we also propose a new observable
based on the angular correlations among hadronic top decay products.
Again, identifying the subjet corresponding to the anti-fermion from W decay,
we construct an angular observable in the top rest frame,
between the direction of a subjet and the boosted tagged top 
momentum in the lab frame.
This angular variable, namely $\cos\theta^\star$(\eqnref{eq:ourobs}) 
turns out to be very
robust to identify the polarization of boosted top quark, 
as shown in \figref{fig:cpsthetastar}.
Evidently, selection of $\cos\theta^\star$ unambiguously discriminate events
due to the left and right handed top quark. Based on this feature we define
an asymmetry(\eqnref{eq:asym}) of event which
was found to be very effective to
measure polarization.
Obviously, this asymmetry is
very sensitive to the
couplings
as shown in \figref{asymwp}. The asymmetry varies within a wide range
from +20\% to -20\% for
$g_{R(L)} =1(0)$ to $g_{L(R)}=1(0)$.   
Understandably, the measurement of asymmetry
is a good indication of the nature of couplings.

The impact of these constructed polarization sensitive variables are studied
for two new physics scenarios where polarized top quarks are produced
from top squark decay in SUSY searches and KK gluon and graviton decay 
in RS model. 
In top squark decay(\eqnref{eq:stopeventtype}), the couplings, and hence the polarization
depend on the details of SUSY parameters. The energy fraction variables,
and angular observable are presented setting certain benchmark parameters
and top squark mass. Remarkably,
the $\cos\theta^\star$ is observed to be very robust.
The variation of polarization 
asymmetry is studied for various input model parameters, 
including mixing angles in stop and neutralino sectors. 
Huge asymmetry is observed ranging from 
+15\% to -25\% depending on the mixing angles in top squark sector. Evidently,
the measurement of asymmetry can shed some insight about the model,
in particular the helicity of top squark and nature of neutralinos, 
which might be
very useful information in reconstructing models.
Similar exercise is also carried out in the context of RS model.
The substantial impact of these constructed variables are observed 
in this model,
and again, the measured asymmetry might be an useful tool to pin down
the nature of couplings in this model.
It is to be noted that there are SM physics processes with
large cross sections which have identical final state
as the processes considered here.
In order to achieve signal sensitivity at a reasonable level,
these backgrounds need to be suppressed by applying kinematic selections.
It is really challenging, and worth to study how those
background optimization cuts affect our observations.
Therefore, one needs to carry out a dedicated analysis
matching the experimental constrains at the LHC,
and establish the robustness of
our strategy to identify the top quark polarization \cite{ourcontinue}.

\section*{Acknowledgments}

AHV is thankful to Tuhin Roy for some useful discussion.
The work of RMG is supported by the Department of Science and Technology,
India under Grant No. SR/S2/JCB-64/2007.
JL and CKK would like to thank Ashwani Kushwaha for many fruitful
discussions related to RS models. CKK wishes to acknowledge support
from the Royal Society-SERB Newton International Fellowship (NF171488).
JL wants to thank TIFR for the hospitality and support during her stay in
TIFR(DHEP), where a part of this work was done.
This work was supported in part by the CNRS LIA-THEP and the INFRE-HEPNET of CEFIPRA/IFCPAR (Indo-French Centre for the Promotion of Advanced Research).

\bibliography{main.bib}
\bibliographystyle{utphys.bst}
\end{document}